\documentclass[lettersize,journal]{IEEEtran}

\def\endthebibliography{%
	\def\@noitemerr{\@latex@warning{Empty `thebibliography' environment}}%
	\endlist
}
\makeatother
\usepackage{amssymb}
\usepackage{amssymb,amsmath,amsthm}
\usepackage[linesnumbered,lined,boxed,commentsnumbered,ruled]{algorithm2e}	
\SetKwRepeat{Do}{do}{while}%

\usepackage{algorithmicx}

\usepackage{stmaryrd}
\SetSymbolFont{stmry}{bold}{U}{stmry}{m}{n}
\usepackage{multirow,url}
\usepackage{color,cite}

\usepackage{graphicx}
\usepackage{float}
\usepackage{subfigure}
\usepackage{overpic}
\usepackage{caption}
\usepackage{bbding}

\usepackage{cite}

\usepackage{tikz,pgf}
\usetikzlibrary{fadings,shapes,arrows,shadows}
\usepackage{makecell}
\usepackage{array}
\usepackage{url}
\usepackage{float}
\usepackage{booktabs} 
\usepackage{tabularx}
\usepackage{amsmath}
\usepackage{amsfonts}
\usepackage{mathrsfs}
\usepackage{amsfonts,amsthm,bm}
\usepackage{cases}

\newtheorem{theorem}{Theorem}

\theoremstyle{plain}

\newcommand{\com}[1]{} 
\renewcommand{\qedsymbol}%
{\rule{1ex}{1.5ex}}

\def\O{\mathcal{O}}
\def\M{\mathcal{M}}
\def\N{\mathcal{N}}
\def\S{\mathcal{S}}
\def\T{\mathcal{T}}

\def\T{\mathcal{T}}
\def\C{C}
\def\L{\mathcal{L}}

\IEEEoverridecommandlockouts
\begin{document}
\title{Joint Edge Server Deployment and Computation Offloading: A Multi-Timescale Stochastic Programming Framework}
\author{Huaizhe Liu, Jiaqi Wu, Zhizongkai Wang, Bin Cao, and Lin Gao \vspace{-5mm}
\thanks{This work was supported in part by the National Key Research and Development Program of China under Grant 2021YFB2900300,     Natural Science Foundation of Guangdong Province under Grant 2024A1515010178,     Shenzhen Science and Technology Program under Grant KQTD20190929172545139, Grant GXWD20231129103946001, Grant KJZD20240903095402004, and Grant ZDSYS20210623091808025, and  Guangdong Basic and Applied Basic Research Foundation under Grant 2023A1515012819. (\emph{Corresponding Authors: Lin Gao, Bin Cao})~~}
\thanks{H. Liu,  Z. Wang, B. Cao, and L. Gao are with the School of Electronics and
Information Engineering and the Guangdong Provincial Key Laboratory of
Aerospace Communication and Networking Technology, Harbin Institute of
Technology, Shenzhen, China. Email: gaol@hit.edu.cn.}
\thanks{J. Wu is with ... Email: ... }
\thanks{Part of the results have been published in IEEE WiOpt 2023 \cite{LHZ-Wiopt-2023}.}
}

\addtolength{\abovedisplayskip}{-1mm}
\addtolength{\belowdisplayskip}{-1mm}

\IEEEtitleabstractindextext{
	
	\begin{abstract}
		\emph{Mobile Edge Computing (MEC)} is a promising approach for enhancing the quality-of-service (QoS) of AI-enabled applications in the B5G/6G era,
		by bringing computation capability closer to end-users at the network edge.
		In this work, we investigate the joint optimization of edge server (ES) deployment, service placement, and computation task offloading under the stochastic information scenario.
		Traditional approaches often treat these decisions as equal, disregarding the differences in information realization.
However, in practice, the ES deployment decision must be made in advance and remain unchanged, prior to the complete realization of information,
		whereas the decisions regarding service placement and computation task offloading  can be made and adjusted in real-time after information is fully realized.
		To address such temporal coupling between decisions and information realization, we introduce the \emph{stochastic programming (SP)} framework, which involves a \emph{strategic-layer} for deciding ES deployment based on (incomplete) stochastic information and a \emph{tactical-layer} for deciding service placement and task offloading based on complete information realization.			The problem is challenging due to the different timescales of two layers' decisions.
		To overcome  this challenge, we propose a multi-timescale SP framework, which includes a large timescale (called period) for strategic-layer decision-making and a small timescale (called slot) for tactical-layer decision making.		
Moreover, we design a Lyapunov-based algorithm to solve the tactical-layer problem at each time slot, and a Markov approximation algorithm to solve the strategic-layer problem in every time period.		
		Simulation results demonstrate that our proposed solution significantly outperforms existing benchmarks that overlook the coupling between decisions and information realization, achieving up to  $56\%$ reduction in total system cost.~~~~
	\end{abstract}
}

\maketitle
\IEEEdisplaynontitleabstractindextext
\IEEEpeerreviewmaketitle

\section{Introduction}\label{section:introduction}
\subsection{Background and Motivations}

{With the rapid advancement of Beyond 5G (B5G) and 6G technologies, a wide array of AI-enabled distributed interactive applications (DIAs) has emerged in everyday life, including collaborative virtual reality (VR), holographic projection, and Metaverse.}
These applications are typically resource-intensive, computationally demanding, and highly sensitive to latency. Traditional cloud computing architectures often fall short in supporting such applications due to significant communication overhead, variable network jitter, and long round-trip times (RTT) between centralized cloud servers and mobile devices \cite{H.Yang_survey}.
To address these limitations, \emph{mobile edge computing (MEC)}  \cite{mec_survey_2022, add-1,add-2,add-3,add-4,add-5}   has emerged as a promising paradigm to support these new applications via offering computation services at the network edge in close proximity to end-users.
Therefore, MEC has been increasingly recognized as a key enabling technology for supporting next-generation applications in the B5G/6G era.

Due to its significant potential in enhancing future networks and applications, MEC has garnered considerable attention from both industry and academia.
Many existing works in this field have mainly focused on the  {computation task offloading} problem and the associated  {resource deployment} problem
\cite{M.Tang_TMC2022, Wujiaqi_ICC2022, T.Liu_TMC2022, Y.Gao_JIOT_2023, H.Song_TNSE2023, Z.Lin_TWC2021, X.Zhang_JIOT2022, Y.Zhang_TON2022, Y.Ma_TMC2022, C.Jiang_TNSE2023, R.Zhou_INFOCOM2022, R.Li_TPDS2022, W.Fan_TMC2023, X.Ma_TMC2023, Y.Shi_TWC2023,X.Dai_TMC2023, Z.Bai_TMC2024, J.Yan_TWC2021, H.Liu_TWC2022, S.Gong_TVT2023, J.Yan_TWC2023}.
However, these studies often \textbf{treat all decisions equally}, without accounting for differences in the level of information available at the time of decision-making.
For example, in \cite{Wujiaqi_ICC2022, Y.Gao_JIOT_2023, H.Song_TNSE2023, Z.Lin_TWC2021, X.Zhang_JIOT2022, Y.Zhang_TON2022, Y.Ma_TMC2022, C.Jiang_TNSE2023}, all decisions are made based on \emph{complete information}, assuming that all relevant information has been realized when making these decisions.
In contrast, in \cite{M.Tang_TMC2022, T.Liu_TMC2022, R.Zhou_INFOCOM2022, R.Li_TPDS2022, W.Fan_TMC2023, X.Ma_TMC2023, Y.Shi_TWC2023, X.Dai_TMC2023, Z.Bai_TMC2024, J.Yan_TWC2021, H.Liu_TWC2022, S.Gong_TVT2023, J.Yan_TWC2023}, all decisions are made based on \emph{incomplete information} (e.g., stochastic distribution information), aiming to optimize the expected or estimated objective function.

In practice, however, \emph{different decisions are often based on different levels of information} due to differences in decision-making timing and the frequency of changes.
For instance, in an MEC network, decisions regarding hardware resource deployment cannot be adjusted too quickly due to the physical limitations of hardware. As a result, such decisions must be made in advance, often before complete information (e.g., user locations, channel conditions, etc.) is fully realized.
In contrast, decisions regarding service placement and computation task offloading can be adjusted more rapidly, with fewer constraints, and therefore can be made in real-time once complete information is realized.
Furthermore,  real-time decisions, such as those related to computation task offloading, are likely to change frequently to accommodate the dynamic delay requirements of real-time applications \cite{J.Santos_TUT,X.Jiang_TUT}.~~~~~~~~~~~~~~~~~~

\begin{figure}
	\centering
	\includegraphics[width=1\linewidth]{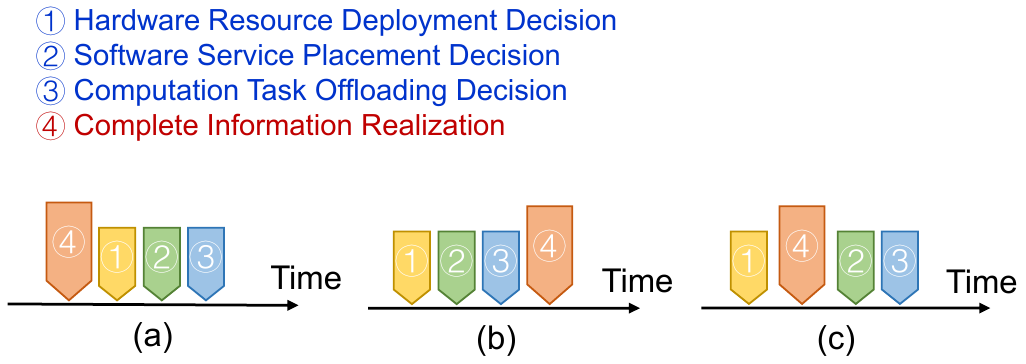}
	\vspace{-2mm}
	\caption{Decision-making based on
	(a) complete information,
	(b) incomplete information,	
		and (c) hybrid information. }
	\label{fig:model}
\end{figure}

In this work, we will consider such \textbf{temporal coupling} between different decisions and information realization.
Specifically, we focus on such a decision-making scheme in which different decisions are made at different times based on different levels of information realized.
We refer to this as decision-making based on \textbf{hybrid information}, in contrast to those based on complete information or incomplete information.
For clarity, we illustrate different decision-making schemes based on different information in Fig.~\ref{fig:model}.
More specifically,
Fig.~\ref{fig:model}(a) illustrates  decision-making based on complete information, where all relevant information is realized before making decisions, and thus all decisions can be made based on complete information.
Fig.~\ref{fig:model}(b) illustrates decision-making based on incomplete information, where no information is available prior to decision-making, and thus all decisions are made using stochastic information.
Fig.~\ref{fig:model}(c) illustrates decision-making based on hybrid information, where the hardware resource deployment decision is made in advance  before complete information is realized, while the software service placement and computation task offloading decisions are made after complete information becomes available. ~~~~~~~~~~~~

Clearly, the first decision-making scheme (based on complete information) assumes the availability of complete information, which may not be feasible in real-world systems.
The second decision-making scheme (based on incomplete information) disregards any available information that might have already been realized, potentially leading to unnecessary
performance losses. The third decision-making scheme (based on hybrid information) proposed in this work optimally utilizes the available information, achieving the best possible performance
under the constraints of limited information. ~~~~~~~~~~~~

\subsection{Solution Approach and Contributions}

\begin{figure}
	\centering
	\includegraphics[width=0.8\linewidth]{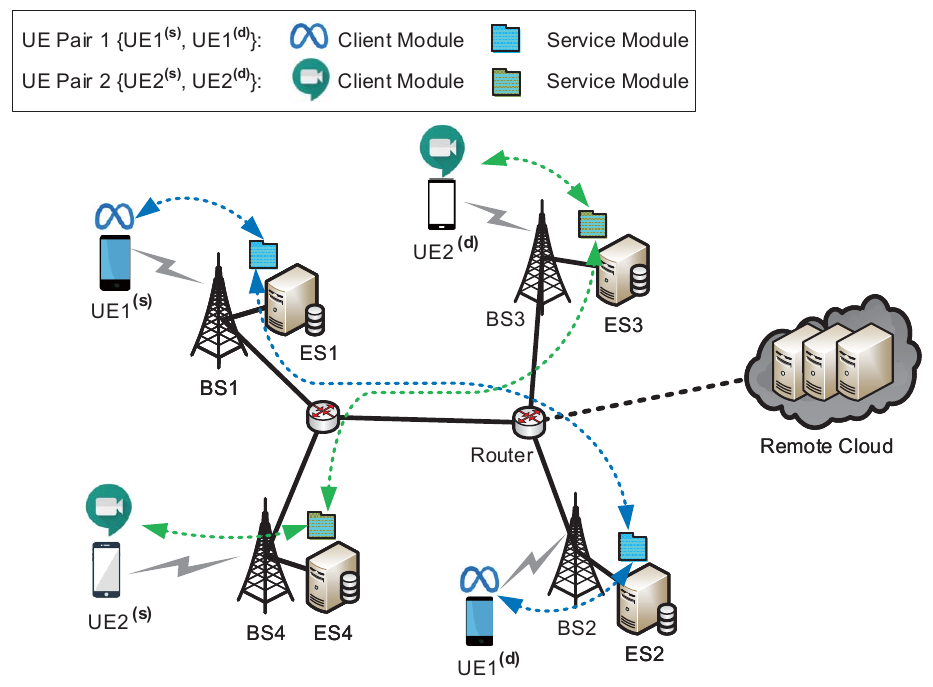}
	\caption{An example of MEC network with DIA applications. }
	\label{fig:system-model}
\end{figure}

In this work, we consider a typical MEC network comprising a set of edge computing servers (ES) deployed at base stations (BS), and a group  of mobile users (UEs) who interact with each other via DIA applications (e.g., collaborative VR) and seek to offload computation tasks to the ES.
As illustrated  in Fig.\ref{fig:system-model}, a DIA application, like many mobile applications, consists of two main modules:
(i) a local module (the \emph{client}) operating on mobile devices, mainly responsible for the \emph{local computation} such as data input/output and preprocessing,
and (ii) a remote module (the \emph{service}) running on the cloud server, mainly responsible for the \emph{core computation} such as image processing and 3D graphics rendering.

The interaction between two UEs follows the following steps:
First, each UE's client module performs the local computation tasks and communicates with its service module on the remote cloud server.
Then, both service modules perform their core computation tasks and communicate with each other for necessary information exchange.
Obviously, in such a scenario, the local computation tasks cannot be offloaded due to the data and operation locality constraints, while the core computation tasks can be offloaded (from the cloud) to an ES, as long as there are sufficient hardware resources and necessary software services in the ES.
In the example of Fig. 2, UE1$^{\rm(s)}$ and UE1$^{\rm(d)}$ offload core computation tasks to ES 1 and ES 2, while UE2$^{\rm{(s)}}$ and UE2$^{\rm{(d)}}$ offload   core computation tasks to ES4 and ES3, respectively.

In such a scenario, we focus on studying the offloading of core computation tasks for UEs.
As noted earlier, core computation tasks of a DIA can be offloaded from the cloud to an ES, \emph{only if} there are sufficient hardware resources and necessary services in the ES.
That is, the core computation offloading decision is highly dependent on the service placement and ES deployment decisions.
To address this, we will study the following problems jointly in this work:
\begin{enumerate}
		\item \textbf{ES Deployment:} Which ES should be deployed at each BS, considering the resource demand uncertainty?
		\item \textbf{Service Placement:} What services should be placed on each ES, taking into account both the service placement cost and the computation offloading requests?
		\item \textbf{Computation Task Offloading:} Which ES should each UE offload its core computation tasks to, given the service placement at each ES?
\end{enumerate}

As mentioned previously, we  consider a  practical scenario where complete information is \emph{not} available upfront, and different decisions can be made based on different levels of information.
Specifically, as illustrated in Fig.~\ref{fig:model} (c), we consider the decision-making scheme based on \emph{hybrid information}, where the ES resource deployment decision must be made well in advance before the complete information is realized,
while decisions regarding service placement and computation task offloading can be made in real time once the complete information is realized.

To capture such temporal coupling between different decisions and information realization, we introduce the \emph{stochastic programming
(SP)} framework, which invoves a \emph{strategic-layer} for making ES deployment decisions, and a \emph{tactical layer} for making service placement and computation task offloading decisions.
The strategic-layer decisions are made based on the (incomplete) stochastic distribution information (i.e., before the complete information is realized), while the tactical-layer decisions are made based the complete information realization.
This problem is very challenging due to the different decision-making times as well as the different timescales of two layers' decisions.
To address these challenges, we propose a \emph{multi-timescale} SP framework, which includes a large timescale (called period) for strategic-layer decision-making and a small timescale (called slot) for tactical-layer decision-making.
To find the optimal decisions, we develop a Lyanpunov-based algorithm to solve the tactical-layer decisions at each time slot, and a Markov approximation algorithm to solve the strategic-layer decisions in every time period.
In summary, the main contributions of this work are   as follows:

\begin{itemize}
	\item \emph{Novel Decision-Making Scheme based on Hybrid Information:}
We propose a novel decision-making scheme based on hybrid information, where different decisions are made based on different levels of information realized.
This scheme does not rely on the complete information assumption, and meanwhile can fully utilize the available information.
Thus, it is more realistic and efficient than those purely based on the complete information or stochastic information.	

	\item \emph{Multi-Timescale Stochastic Programming Formulation:}
To capture the temporal coupling between decisions and information realization, we propose a multi-timescale SP framework, which includes a large-timescale strategic layer for making long-term decisions and a small-timescale tactical layer for making short-term decisions.
We further develop a Lyapunov-based algorithm to solve the tactical-layer decisions at each time slot, and a Markov approximation algorithm to solve the strategic-layer decisions in every time period.

	\item \emph{Performance Evaluations and Insights:}
	Simulation results demonstrate that our proposed solution significantly outperforms existing benchmarks, achieving up to a $56\%$ reduction in the total system cost.
This implies that our proposed decision-making scheme based on hybrid information can better utilize the available information, achieving better possible performance under the constraints of limited information.


\end{itemize}

The rest of the paper is organized as follows.
In Section \ref{section:related_works}, we review related works.
In Section \ref{section:model}, we present system model.
In Section \ref{section:problem_formulation}, we formulate the multi-timescale SP problem.
In Section \ref{section:problem_transformation_and_solution}, we present the detailed algorithms to solve the SP problem.
In Section \ref{section:simulation_results}, we present the simulation results.
Finally, we conclude in Section \ref{section:conclusion}.

\section{Related Works}\label{section:related_works}

\begin{table}[t]
	\begin{center}\caption{A comparison with existing works}\label{reference_compare}
		\vspace{-2mm}
		{\fontsize{7.2}{7.2}\selectfont
			\begin{tabular}{|c|c|c|c|c|}
				\hline
				\makecell[c]{References} & \makecell[c]{ES \\ deployment} & \makecell[c]{Service \\ placement} & \makecell[c]{Task \\ Offloading} & \makecell[c]{Information}     \\
				\hline
				\cite{Wujiaqi_ICC2022} &  &  & \checkmark & \multirow{5}{*}{Complete}    \\
				\cline{1-4}
				\cite{Y.Gao_JIOT_2023,H.Song_TNSE2023} & \checkmark  &  & \checkmark &     \\
				\cline{1-4}
				\cite{Z.Lin_TWC2021,Y.Ma_TMC2022} &  & \checkmark & \checkmark &       \\
				\cline{1-4}
				\cite{Y.Zhang_TON2022} & \checkmark &  &  &     \\
				\cline{1-4}
				\cite{X.Zhang_JIOT2022,C.Jiang_TNSE2023} & \checkmark & \checkmark &  &       \\
				\hline
				\cite{T.Liu_TMC2022,M.Tang_TMC2022,X.Ma_TMC2023,X.Dai_TMC2023,Z.Bai_TMC2024} &  &  & \checkmark &\multirow{6}{*}{Stochastic} \\
				\cline{1-4}
				\cite{W.Fan_TMC2023} & \checkmark & \checkmark & \checkmark & \\
				\cline{1-4}
				\cite{R.Zhou_INFOCOM2022,Y.Shi_TWC2023} &  & \checkmark & \checkmark &      \\
				\cline{1-4}
				\cite{R.Li_TPDS2022} &  & \checkmark &  &        \\
				\cline{1-4}
				 \cite{H.Liu_TWC2022, S.Gong_TVT2023}& & & \checkmark &   \\
				\cline{1-4}
				\cite{J.Yan_TWC2021,J.Yan_TWC2023} &  & \checkmark & \checkmark &   \\
				\hline
				\textbf{Our work} & \checkmark & \checkmark & \checkmark & \textbf{Hybrid}  \\
				\hline
			\end{tabular}
		}
	\end{center}
\end{table}

Recently, MEC has garnered considerable attention from both industry and academia.
Many existing works in this field have mainly focused on the joint optimization of {computation task offloading} and the associated service placement and {resource deployment},  based on either complete information \cite{Wujiaqi_ICC2022, Y.Gao_JIOT_2023, H.Song_TNSE2023, Z.Lin_TWC2021, X.Zhang_JIOT2022, Y.Zhang_TON2022, Y.Ma_TMC2022, C.Jiang_TNSE2023} or incomplete stochastic information \cite{M.Tang_TMC2022, T.Liu_TMC2022, R.Zhou_INFOCOM2022, R.Li_TPDS2022, W.Fan_TMC2023, X.Ma_TMC2023, Y.Shi_TWC2023, X.Dai_TMC2023, Z.Bai_TMC2024, J.Yan_TWC2021, H.Liu_TWC2022, S.Gong_TVT2023, J.Yan_TWC2023}.

%

\subsection{Joint Optimization under Complete Information}

In \cite{Wujiaqi_ICC2022, Y.Gao_JIOT_2023, H.Song_TNSE2023, Z.Lin_TWC2021, X.Zhang_JIOT2022, Y.Zhang_TON2022, Y.Ma_TMC2022, C.Jiang_TNSE2023},
all decisions are made based on \emph{complete information}, assuming that all relevant information has been realized when making these decisions.
For example,
Wu \emph{et al.} in \cite{Wujiaqi_ICC2022} studied a joint computation offloading and resource allocation problem with the aim of minimizing task delay, considering the task queueing dynamic.
Gao \emph{et al.} in \cite{Y.Gao_JIOT_2023} proposed a two-step optimization framework aimed at minimizing the ES deployment and task scheduling costs.
Song \emph{et al.} in \cite{H.Song_TNSE2023} investegated the joint ES deployment and user offloading association problem and designed a heuristics solution based on genetic algorithm.
Lin \emph{et al.} in \cite{Z.Lin_TWC2021} considered the AI-service placement problem in a multi-user MEC system. By jointly optimizing the service placement and resource allocation, the authors aimed to minimize the total computation time and energy consumption of all users.
Taking into account user mobility and service delay requirements, Ma \emph{et al.} in \cite{Y.Ma_TMC2022} optimized the admission of user service requests with the goal of maximizing both accumulative network utility and accumulative network throughput.
Zhang \emph{et al.} in \cite{Y.Zhang_TON2022} addressed an MEC server deployment problem for network operators, formulating it as a revenue maximization challenge with the goal of ensuring the profitability of the MEC system.
To maximize the overall profit of all ES, Zhang \emph{et al.} in \cite{X.Zhang_JIOT2022} jointly considered the ES deployment and service placement problem in MEC network, aiming to maximize the overall profit of all ES.
To enhance the utility of network participants, Jiang \emph{et al.} in \cite{C.Jiang_TNSE2023} proposed a peer service caching framework. In this framework, edge devices can choose to cache services and share the cached contents with other peer devices.

Obviously, the decision-making scheme in \cite{Wujiaqi_ICC2022, Y.Gao_JIOT_2023, H.Song_TNSE2023, Z.Lin_TWC2021, X.Zhang_JIOT2022, Y.Zhang_TON2022, Y.Ma_TMC2022, C.Jiang_TNSE2023} relies on the assumption of complete information.
However, this assumption may not be practical in real-world systems, especially in the fast-changing environments.

\subsection{Joint Optimization under Stochastic Information}

In contrast, in \cite{M.Tang_TMC2022, T.Liu_TMC2022, R.Zhou_INFOCOM2022, R.Li_TPDS2022, W.Fan_TMC2023, X.Ma_TMC2023, Y.Shi_TWC2023, X.Dai_TMC2023, Z.Bai_TMC2024, J.Yan_TWC2021, H.Liu_TWC2022, S.Gong_TVT2023, J.Yan_TWC2023}, all decisions are made based on \emph{incomplete information} (e.g., stochastic distribution information), aiming to optimize expected or estimated objective.

Specifically, in \cite{M.Tang_TMC2022, T.Liu_TMC2022, R.Zhou_INFOCOM2022, R.Li_TPDS2022, W.Fan_TMC2023, X.Ma_TMC2023, Y.Shi_TWC2023, X.Dai_TMC2023, Z.Bai_TMC2024}, authors formulated long-term optimization problems to optimize expected objectives, and developed online algorithm based on the Lyapunov framework.
For example,
Liu \emph{et al.} in \cite{T.Liu_TMC2022} proposed an online task offloading and resource allocation approach for edge-cloud orchestrated computing, aiming at minimize the average latency of tasks over time.
Tang \emph{et al.} in \cite{M.Tang_TMC2022} proposed an online learning-based distributed offloading algorithm, where each device determines its offloading decision based on the current environment.
The works \cite{X.Dai_TMC2023} and \cite{Z.Bai_TMC2024} studied a UAV-assisted task offloading problem, aimed at alleviating task processing delay.
Ma \emph{et al.} in \cite{X.Ma_TMC2023} considered a dynamic task offloading problem in a cloud-assisted edge computing scenario, aiming to minimize average task response time within resource budget limits.
Fan \emph{et al.} in \cite{W.Fan_TMC2023} proposed a scheme for joint service placement, task scheduling, computing resource allocation, and transmission rate allocation in a multi-task and multi-service scenario with edge-cloud cooperation scheme. The scheme aims to minimize the total task processing delay while ensuring long-term stability in task queuing.
Shi \emph{et al.} in \cite{Y.Shi_TWC2023} proposed a two-timescale optimization framework to jointly optimize task rerouting, service migration, and resource allocation decisions, considering users' mobility.
Zhou \emph{et al.} in \cite{R.Zhou_INFOCOM2022} proposed a decentralized optimization framework to minimize task latency while considering the unique features of both on-ground users and UAVs.
Li \emph{et al.} in \cite{R.Li_TPDS2022} considered a joint optimization problem for application placement and request routing to maximize system performance, subject to a long-term budget constraint on application reconfiguration costs.
However, the above works aim to optimize system performance in a single stage based on stochastic information, ignoring  the  long-term infrastructure planning from dynamic runtime adjustments.

Different from \cite{M.Tang_TMC2022, T.Liu_TMC2022, R.Zhou_INFOCOM2022, R.Li_TPDS2022, W.Fan_TMC2023, X.Ma_TMC2023, Y.Shi_TWC2023, X.Dai_TMC2023, Z.Bai_TMC2024},
researchers in \cite{J.Yan_TWC2021, H.Liu_TWC2022, S.Gong_TVT2023, J.Yan_TWC2023}
developed methods to optimize estimated objective derived from distribution functions.
For example,
Yan \emph{et al.} in \cite{J.Yan_TWC2021} proposed an MEC service pricing scheme to coordinate service caching decisions and control users' task offloading behaviors.
Liu \emph{et al.} in \cite{H.Liu_TWC2022} studied a computation offloading problem for video analytics, aiming to maximize both accuracy and frame processing rate.
Gong \emph{et al.} in \cite{S.Gong_TVT2023} proposed a framework to optimize computation task offloading for ground users (GUs) using multiple UAVs coordinated by a base station, with the goal of minimizing UAVs' trajectories while balancing energy consumption and data queue size across the UAVs.
Yan \emph{et al.} in \cite{J.Yan_TWC2023} proposed a joint computation task offloading and MEC resource deployment scheme based on a Bayesian optimization framework, with the aim of minimizing the accumulated energy-delay cost for users under dynamic system states.

However, decisions in \cite{M.Tang_TMC2022, T.Liu_TMC2022, R.Zhou_INFOCOM2022, R.Li_TPDS2022, W.Fan_TMC2023, X.Ma_TMC2023, Y.Shi_TWC2023, X.Dai_TMC2023, Z.Bai_TMC2024, J.Yan_TWC2021, H.Liu_TWC2022, S.Gong_TVT2023, J.Yan_TWC2023} disregards any available information that might
have already been realized, potentially leading to unnecessary
performance losses.

\subsection{Novelty of Our Work}

In summary, the above existing works treat all decisions equally,
without accounting for differences in the level of information
available at the time of decision-making.
In this work, we will consider the \emph{temporal coupling} between different decisions and information realization.
More specifically, we propose the decision-making scheme based on \emph{hybrid information}, in which different decisions are made at different times based on different levels of information realized.
Such a novel decision-making scheme can optimally utilize
the available information, achieving the best possible performance under the constraints of limited information.

For clarity, we summarize the key features of our work and the existing works in Table \ref{section:related_works}.
%


\section{System Model}\label{section:model}

\subsection{Network Model}\label{section:model:network}
As shown in Fig.~\ref{fig:system-model}, we consider an MEC network consisting of a set $\mathcal{M}=\{1,2,\cdots,|\M|\}$ of $|\M|$ base stations (BS) and a set $\mathcal{N}=\{1,2,\cdots,|\N|\}$ of $|\N|$ mobile UE pairs.
Each BS $m$ can deploy an ES, also denoted by $m$ for notational convenience.
Let $\C_m$ (in GHz) and $\Phi_{m}$ (in GB) denote the available computation capacity and storage capacity on ES $m$,  respectively.
Each UE pair $n $  consists of a source UE (denoted by $\mathrm{UE}_n^{\rm(s)}$) and a destination UE (denoted by $\mathrm{UE}_n^{\rm(d)}$), who interact with each other via a particular DIA.
Different UE pairs may run different DIAs, e.g., UE pair 1 runs collaborative VR (e.g., Meta Horizon Workrooms) and UE pair 2 runs holographic projection (e.g., Google Meet) in Fig.~\ref{fig:system-model}.

Similar as in many existing literatures, we consider a \emph{time-slotted} system, where each time period (e.g., one day or one month) is devided into a set $\mathcal{T} = \{1,2,\cdots, |\mathcal{T}| \}$ of $|\mathcal{T}|$  time slots with the same time length (e.g., 5 seconds).
Moreover, we consider a \emph{quasi-static} system, where the network state (e.g., UE locations, wireless channel environments, and available resources) remains unchanged within each time slot, but may change across different time slots. For convenience, we list the key notations used in this work in Table \ref{main_notations}.

\begin{table}[t]
	\caption{Main notations in this work}
	\label{main_notations}
\hspace{-3mm}
	\begin{tabular}{c|p{25em}}
		\hline
		\textbf{Symbols} & \textbf{Description} \\
		\hline
		\hline
		$\mathcal{M}$ & The sets of BS (ES), $m \in \mathcal{M}$ denoting a particular BS (ES)\\
		$\mathcal{N}$ & The set of UE pairs, $n \in \mathcal{N}$ denoting a particular UE pair \\
		$\mathcal{S}$ & The set of services, $s \in  \mathcal{S}$ denoting a particular service \\
		$\mathcal{T}$& The set of time slots, $t\in \mathcal{T}$ denoting a particular time slot \\
		$\mathrm{i} \in \{\mathrm{s},\mathrm{d} \}$ & The source and destination of a UE pair \\
		$R^{(\mathrm{i})}_{n,m}(t)$ & The channel capacity between the source (or destination) of UE pair $n$ and BS $m$ at time slot $t$ \\
		$T^{(\mathrm{i})}_{n,m}(t)$ & The transmission delay cost of the source (or destination) of UE pair $n$ to BS $m$ at time slot $t$\\
		$D_{n,m}^{(\rm{i})}(t)$		& The computation delay cost of the source (or destination) of UE pair $n$	on ES $m$ at time slot $t$ \\
		$C_{m}(t)$	  &	 The available computation resource on ES $m$ at time  slot $t$ \\
		$\Phi_{m}(t)$ &  The available storage resource on   ES $m$  at time slot $t$  \\
		$\omega(t)$   & The network information realized at time slot $t$, i.e., $\omega(t) \triangleq \{ \Phi_m(t), \C_m(t) , \forall m\in  \mathcal{M} \}$ \\
		$\boldsymbol{\omega}  $ & The complete network information realized at all time slots, i.e., $\boldsymbol{\omega} \triangleq (\omega(t), t\in \T) $\\
		$P^{idle}_{m}$ & The idle CPU power consumption on ES $m$ \\
		$P^{max}_{m}$ & The maximum CPU power consumption on ES $m$ \\
		$P_{avg}$ & The long-term energy budget of all ES in the network \\
		$f_{n}$ & The interaction frequency between the source and destination of UE pair $n$ \\
		\hline
	\end{tabular}
\end{table}

\subsection{Application and Service Model}\label{section:model:service}
We consider a simple but representative DIA model, as shown in Fig.2, where each DIA consists of two components: a local module (called \emph{client}) responsible for the local computation and a remote module (called \emph{service}) responsible for the core computation.
To reduce latency and improve QoS, an UE can offload the core computation tasks to a nearby ES (from the cloud server), if the hardware resources are sufficient and the corresponding service is available on the ES.
In the example of Fig.2,
UE1$^{\rm{(s)}}$, and UE1$^{\rm{(d)}}$ offload tasks to ES1 and ES2, respectively, where the corresponding service module for Metaverse has been placed on ES1 and ES2. Similarly, UE2$^{\rm{(s)}}$, and UE2$^{\rm{(d)}}$ offload tasks to ES4 and ES3, respectively, where the corresponding service module for Google Meet has been placed on ES4 and ES3.

Let $\mathcal{S}=\{1,2,\cdots,|\S|\}$ denote the set of all services, $s_n \in \mathcal{S} $ denote the service of the $n$-th UE pair, and $f_n$ denote the interaction frequency between the source and destination of the $n$-th UE pair.\footnote{{For simplicity, we adopt the \emph{normalized} interaction frequency in this study, where typical interaction frequency values are scaled between 0 and 1. In our simulations, we choose a moderate value (i.e, $f_n = 0.5$) as a representative example  to demonstrate the general trends in the results.}}
Based on the above, we can characterize each service $s \in \mathcal{S} $ by a 5-tuple:
\begin{equation}
	<u_{s},b_{s},c_{s},d_{s},e_{s}>,
\end{equation}
 where $u_{s}$ (in GB) denotes the storage size required by service $s$, $b_{s}$ (in GHz) denotes the local computation workload (of the client module) at UE devices, $c_{s}$ (in GHz) denotes the core computation workload (of the service module) at the cloud server, $d_{s}$  (in MB) denotes the local data volume transmitted between the client and service modules, $e_{s}$ (in MB) denotes the remote data volume exchanged between two service modules of source and destination.

 For notational convenience, we further introduce notations  $u_n \triangleq u_{s_n}$, $ b_n\triangleq b_{s_n}$, $c_n\triangleq c_{s_n}$, $d_n \triangleq d_{s_n}$, and $e_n\triangleq e_{s_n}$ to denote the service storage size, local computation task load, core computation task load, local data volume, and remote data volume corresponding to the DIA runned by UE pair $n$, respectively.
It is worth noting that only the core computation task can be offloaded (from the cloud server to ES), while the local computation task cannot be offloaded due to the data and operation locality of the client module.
Thus, for convenience, we normalize the local computation task load $b_{n}$ to zero.

\subsection{Computation Model}\label{section:model:Computation}
As mentioned before, the available computation resource on each ES $m$ is denoted by $\C_m$ (in GHz). Thus, the time for executing one unit of computation task is $1/\C_m$, and the corresponding computation delay cost is defined as $\lambda_m = \alpha / \C_m $, where $\alpha$ is a frequency-dependent coefficient evaluating the delay cost of one unit of computation time.
Let $D_{n,m}^{(\rm{s})}(t)$ (or $D_{n,m}^{(\rm{d})}(t)$) denote the computation delay costs of the source (or destination) of UE pair $n$ when offloading core computation tasks to ES $m$ at time slot $t\in \T$.
Then,
	\begin{equation}\small \label{equation:computation model src}
		D_{n,m}^{(\rm{i})}(t) = c_{n}\cdot \lambda_{m},~~ \forall m\in \mathcal{M}, n\in \mathcal{N}, \rm{i} \in \rm\{s,d\},
	\end{equation}
where $c_n  = c_{s_n}$ denotes the core computation workload of each user in the UE pair $n$, and $\rm{i} \in \rm\{s,d\}$ denotes the source UE and the destination UE, respectively.

\subsection{Communication Model}\label{section:model:Communication }
As shown in Fig. 2, when UEs offload their core computation tasks to ES (on BS), they need to transmit some local data to the respective service modules in the ES and both service modules need to exchange some remote data. Note that the local data transmission is often achieved via wireless communication links between UEs and BS (ES), and the remote data exchange is often achieved via wired communication links (e.g., optical fiber) between BS (ES). Next, we characterize these two communication links.

\subsubsection{Communication links between UEs and ES}
Each ES $m$ is deployed on BS $m$. Thus, an UE will communicate with an ES via the radio access links of the corresponding BS. Let $R_{n,m}^{(\rm{s})}(t)$ (or $R_{n,m}^{(\rm{d})}(t)$) denote the channel capacity of the radio access link between the source (or destination) of UE pair $n$ and the BS $m$ at time slot $t\in \T$.
According to Shannon-Hartley theorem, we have:
\begin{equation}\small \label{equation:transmission-rate-for-src}
	\begin{aligned}
		\textstyle
		{R_{n,m}^{(\rm{i})}}(t) = {W}\cdot{\log}\left( {1 + {{p_{n,m}^{(\rm{i})}(t) \cdot {h_{n,m}^{(\rm{i})}}}(t) \over {{N_0}}}} \right), \\ \forall m\in \mathcal{M}, n\in \mathcal{N}, \rm \rm{i} \in\{s,d\},
	\end{aligned}
	\textstyle
\end{equation}
where $ p_{n,m}(t)$ is the transmission power between UE   $n$ and BS $m$,
$ h_{n,m}^{(\rm{i})}(t)$ is the channel gain between UE   $n$ and BS $m$,  $W$ is the channel bandwidth, and $N_0$ is the Gauss noise power.
We further set the channel gain $h_{n,m}^{(\rm{i})}(t)= d_{n,m}^{-\theta}$, where $d_{n,m}$ is the distance between UE $n$ and BS $m$,
and $\theta \in [2,4]$ is the path-loss factor.

Let $T_{n,m}^{(\rm{s})}(t)$ (or $T_{n,m}^{(\rm{d})}(t)$) denote the transmission delay costs of the source (or destination) of UE $n$ when transmitting local data to ES $m$ at time slot $t\in \T$.
Then, we have:
	\begin{equation} \small \label{equation:transmission-time-for-src}
		\begin{split}
		{T_{n,m}^{(\rm{i})}(t)} = \beta \cdot \tfrac{d_{n}}{R_{n,m}^{(\rm{i})}(t)}, \forall m\in \mathcal{M}, n\in \mathcal{N}, \rm{i} \in \{\rm{s,d}\},
		\end{split}
	\end{equation}
where $d_n = d_{s_n}$ is the total data transmission volume between the client module  and the service module, and $\beta$ is a coefficient for evaluating the cost of one unit of transmission time.

\subsubsection{Communiation links between ES}
BS are connected with each other through wired links (e.g., optical fiber). Let $R_{m_1,m_2}(t)$ denote the wired link capacity between BS $m_1$ and BS $m_2$ at time slot $t \in \T$. Then, the cost of exchanging remote date of the UE pair $n$ between BS $m_1$ and $m_2$ at time slot $t \in \T$, denoted by $T_{n,m_1,m_2}^{(\rm{e})} (t)$, can be written as:
	\begin{equation} \small \label{equation:transmission-rate-for-des}
		\begin{aligned}
		T_{n,m_1,m_2}^{(\rm{e})} (t) = \beta \cdot \tfrac{e_{n}}{R_{{m_{1},m_{2}}}(t)},  \forall m_1, m_2\in \mathcal{M}, n\in \mathcal{N},
		\end{aligned}
	\end{equation}
where  $e_n = e_{s_n}$ is the total data exchange volume between two service modules. For notational convenience, we set $R_{m,m}(t) = \infty$, and thus $T_{n,m,m}(t) = 0$. That is, when a UE pair $n$ offload core computation tasks to the same ES $m$, the cost of exchanging remote data (within the ES) is zero.

\subsection{Stochastic Information Model}
As mentioned in Section \ref{section:model:network}, we consider a quasi-static system, where the network state (e.g., UE locations, wireless channel environments, and available resources) change randomly across different time slots.
Without loss of generality, we define the \emph{network information} at time slot $t \in \T$, denoted by $\omega(t)$, as the available storage capacity and computation capacity of all ES $m \in \mathcal{M}$ at time slot $t$, i.e.,
\begin{equation}
\omega(t) \triangleq \{ \Phi_m(t), \C_m(t) , \forall m\in  \mathcal{M} \}
\end{equation}
We further introduction notation $\boldsymbol{\omega}$ to denote the complete information at all time slots $\mathcal{T}$, i.e., $ \boldsymbol{\omega} \triangleq (\omega(t), t\in \T)$.

As mentioned previously, we consider a practical (incomplete) \emph{stochastic information scenario}, where the complete information $\boldsymbol{\omega} $ is not available upfront. That is, at the beginning of the time period, we cannot obtain the detailed values of $\omega(t), \forall t \in \mathcal{T}$.
In contrast, the detailed value of $\omega(t)$ is realized at time slot $t$, and thus can only be observed at time slot $t$.
As in many existing literatures, we assume that the stochastic distribution information of $\boldsymbol{\omega} $ is known in advance, e.g., the probability distribution function $f_{\boldsymbol{\omega}}(\boldsymbol{\omega})$.

\section{Problem Formulation}\label{section:problem_formulation}

\begin{figure}
	\centering
	\includegraphics[width=0.9\linewidth]{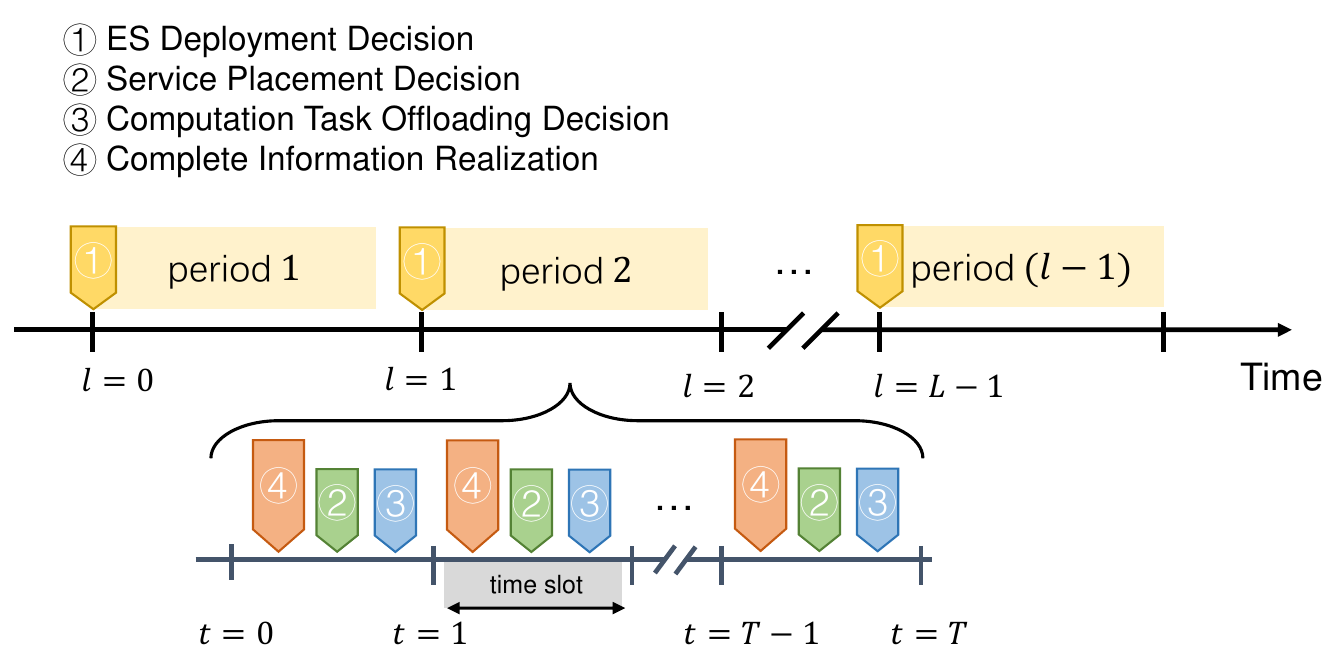}
	\vspace{-2mm}
	\caption{Decision-Making based on hybrid information in multi-period and multi-slot scenario.}
	\label{fig:decision_sequence}
	\vspace{-3mm}
\end{figure}

\subsection{Decision Variables and Key Constraints}
By extending the decision-making scheme in Fig.~\ref{fig:model}(c) into the multi-period and multi-slot scenario, we can obtain  Fig.~\ref{fig:decision_sequence}, which illustrates the decision-making and information realization sequences in $L$ time periods, each consisting of $T$ time slots.
In each time period,
the ES deployment decision is made only once at the beginning of the time period and keeps unchanged in the entire time period, prior
to the realization of complete information in this period.
In contrast, the service placement and computation task offloading decisions are made at each time slot, based on the complete realization of network information at that time slot.

Next, we define the decision variables regarding ES deployment, service placement, and computation offloading.
{Without loss of generality, we consider a particular time period $l$, and omit the time period index $l$ for notational convenience.}

\subsubsection{Decision Variables}
~

\noindent
(i) \emph{ES deployment}: $z_{m} \in \{0,1\}$, which indicates whether an ES $m$ is deployed in the time period;\\
(ii) \emph{Service placement}: $x_{m,s}(t) \in \{0, 1\} $, which indicates whether a service $s$ is placed on ES $m$ at time slot $t\in  \T$;\\
(iii) \emph{Computation offloading}: $y_{m,n}^{(\mathrm{s})}(t),y_{m,n}^{(\mathrm{d})}(t) \in \{0, 1\} $, which indicates whether a source or destination of UE pair $n$ offloads its core computation task to ES $m$ at time slot $t\in \T$.

For convenience, we further introduce the following notations to denote all ES deployment decisions for the time period, and all service placement and computation offloading decisions for all time slots $\forall t \in \T$ in the time period:
\begin{equation}\small\label{decision_variables}
	\begin{aligned}
		&\mathbf{z}\triangleq ( z_m , \forall m \in \mathcal{M}),\\
		&\mathbf{x}(t)\triangleq ( x_{m,s}(t)  , \forall m \in \mathcal{M}, s \in \mathcal{S} ),~~ \forall t \in \T, \\
		&\mathbf{y}(t)\triangleq ( y_{m,n}^{(\mathrm{s})}(t),y_{m,n}^{(\mathrm{d})}(t)  , \forall m \in \mathcal{M}, n \in \mathcal{S}),~~ \forall t \in \T.
	\end{aligned}
\end{equation}

\subsubsection{Key Constraints}
Based on the above decision variables, we now characterize the key constraints.
First, each UE can only connect to one ES at each time slot. Thus, for each UE, we have the following UE connection constraint:
\begin{align} \small \label{UE_connection_constraint}
		\sum_{m\in \mathcal{M}}y_{m,n}^{(\rm{i})}(t)=1, \forall n\in \mathcal{N}, \rm{i} \in \rm{\{s, d\}}, \mathit{t} \in \T.
\end{align}

Second, each service will occupy certain storage resource of the ES that it is placed. Thus, for each ES $m$, we have the following ES storage resource constraint:
\begin{align} \small \label{ES_storage_constraint}
		\sum_{s\in \mathcal{S}}x_{m,s}(t) \cdot u_{s} \leq \Phi_{m}(t), \forall m\in \mathcal{M}, \mathit{t} \in \T.
\end{align}

Third, each UE will occupy certain computation resource of the ES that it offloads to.
Thus, for each ES $m$, we have the following ES computation resource constraint:
\begin{equation} \small \label{ES_computation_constraint}
	\begin{aligned}
	\sum_{n\in \mathcal{N}} f_{n} \cdot  (y_{m,n}^{(\rm{s})}(t) +  y_{m,n}^{(\rm{d})}(t)) \cdot c_{n} \leq \C_m(t), \\  \forall m\in \mathcal{M}, \rm{i} \in \rm{\{s, d\}}, \mathit{t} \in \T .
	\end{aligned}
\end{equation}
where $f_{n}$ is the interaction frequency of UEs in the pair $n$.

Fourth, an UE can offload its core computation task to an ES only if the corresponding service is placed on the ES.
Thus, we have the following service availability constraint:
\begin{equation} \small \label{service_constraint}
	\begin{aligned}
		y_{m,n}^{(\rm{i})}(t) \leq {x_{m,s_{n}}(t)},  \forall m\in \mathcal{M}, n\in \mathcal{N},  \rm{i} \in \rm{\{s, d\}}, \mathit{t} \in \T .
	\end{aligned}
\end{equation}

Fifth, services can be placed on an ES only if the corresponding ES is deployed. Thus, we have the following server availability constraint:
\begin{align} \small \label{ES_constraint}
	\begin{aligned}
		x_{m,s,}(t) \leq {z_m},  \forall m\in \mathcal{M}, s\in \mathcal{S}, t\in \T.
	\end{aligned}
\end{align}

Sixth, there is a limit budget for ES deployment.
Thus, we have the following ES deployment budget constraint:
\begin{equation} \small \label{ES_deploy_constraint}
	\begin{aligned}
		\sum_{m \in \mathcal{M}} q_m \cdot z_m& \leq C_{tot},
	\end{aligned}
\end{equation}
where $q_{m}$ denotes the unit deployement cost of an ES, and $C_{tot}$ denotes the total ES deployment budget.

Last, the ES energy consumption shall satisfy a long-term energy budget.
According to \cite{ES_energy2019}, when placing services on an ES, their operations require CPU resources, which directly affects the ES's overall power consumption. Furthermore, such an energy consumption is closely related to the CPU utilization ratio.
Let $\zeta(t)$ denote the total energy consumption of all ES at time slot $t\in \T$. Then,
\begin{equation}  \label{ES_energy_constraint}
	\begin{aligned}\small
		\displaystyle
		\zeta(t)=\sum_{m \in \mathcal{M}} \bigg(P^{idle}_m + (P^{max}_m - P^{idle}_m) \cdot {\sum_{s \in \mathcal{S}}x_{m,s}(t) \cdot c_{s} \over C_m(t)} \bigg),
	\end{aligned}
\end{equation}
where $P^{max}_m$ denotes the CPU power consumption when ES $m\in \mathcal{M}$ is under the full workload, and $P^{idle}_m$ denotes the CPU power consumption when ES $m\in \mathcal{M}$ is idle.
Thus, we can define the long-term ES energy constraint as follows:
\begin{align}  \label{time_avg_constraint}
	\begin{aligned}\small
		\lim_{|\T|\rightarrow\infty} {1 \over |\T|}\sum\limits_{t=0}^{|\T|-1}\zeta(t) \leq P_{avg},
	\end{aligned}
\end{align}
where $P_{avg}$ denotes the long-term energy budget of all ES, and $|\T| = T$ denote the total time slots in the time period.

\subsection{Optimization Problem Formulation}
Now we define the ES deployment cost, the service placement cost, the UE delay cost, and the total system cost.

\subsubsection{ES Deployment Cost}
To provide computation and storage services to UEs, ES deployment is essential, which inevitably incurs deployment costs.
 Let $q_{m}$ denote the unit ES deployment cost, i.e., the cost of deploying an ES $m$.
Let $\Gamma^{\rm{D}}$ denote the total ES deployment cost. That is,
\begin{equation} \small \label{ES_deployment_cost}
	\begin{aligned}
		\Gamma^{\rm{D}}( \mathbf{z} ) = \sum_{m\in \mathcal{M}} q_{m}\cdot z_{m}.
	\end{aligned}
\end{equation}

\subsubsection{Service Maintenance Cost}
An ES needs to perform routine maintenance for all services placed on it to ensure the stable and smooth running of services, which, inevitably introduces a service maintenance cost.
Let $\rho_{m}$ denote the unit cost of maintaining service on ES $m$.
Let $\Gamma^{\rm{sm}}(\mathbf{x}(t)|\omega(t)) $ denote the total service maintenance cost under network information $\omega(t)$ at time slot $t$.
That is,
\begin{equation} \small \label{service_maintenance_cost}
	\begin{aligned}
		&\Gamma^{\rm{sm}}(\mathbf{x}(t)|\omega(t))  = \sum_{m\in \mathcal{M}}\sum_{s\in \mathcal{S}}\rho_{m}\cdot  x_{m,s}(t) \cdot u_{s}, \forall t\in \T.
	\end{aligned}
\end{equation}

\subsubsection{Service Placement Cost}
In order to place a service within an ES, it is necessary to download and set up the service, which will inevitably incur a cost for service placement.
Let $\theta_{m}$ denote the unit cost of placing service on ES $m$.
Let $\Gamma^{\rm{sp}}(\mathbf{x}(t)|\omega(t)) $ denote the total service placement cost under network information $\omega(t)$ at time slot $t$.
That is,
\begin{equation} \small \label{service_placement_cost}
	\begin{aligned}
		&\Gamma^{\rm{sp}}(\mathbf{x}(t)|\omega(t)) & \\ & = \sum_{m\in \mathcal{M}}\sum_{s\in \mathcal{S}}\theta_{m}\cdot [x_{m,s}(t)-x_{m,s}(t-1)]^{+} \cdot u_{s}, \forall t\in \T,
	\end{aligned}
\end{equation}
where $[a]^{+}=\max\{0,a\}$.
For notational convenience, we denote $\Gamma^{\rm{sop}}(\mathbf{x}(t)|\omega(t))$ as the service operation cost, which consists of both maintenance and placement cost. That is,
\begin{equation} \small \label{service_operation_cost}
	\begin{aligned}
		&\Gamma^{\rm{sop}}(\mathbf{x}(t)|\omega(t)) = \Gamma^{\rm{sm}}(\mathbf{x}(t)|\omega(t)) + \Gamma^{\rm{sp}}(\mathbf{x}(t)|\omega(t)), \forall t\in \T.
	\end{aligned}
\end{equation}

\subsubsection{UE Delay Cost}
The QoS of UE is greatly affected by the interaction latency, i.e., the end-to-end delay between two interacting UEs.
As shown in Fig.~\ref{fig:system-model}, when a pair of interacting UEs offload tasks to ES, the end-to-end delay consists of the following components: (i) the communication and computation delay between the source and the target BS (ES),
(ii) the communication and computation delay between the destination and the target BS (ES), and (iii) the communication delay between two target ES.
Furthermore, due to   resource limitations on ES, some UEs may be unable to offload their tasks to the ES and   need to connect to the remote cloud server as usual.
In this work, we we simplify this cloud computing process by introducing a parameter $T_{\rm{cld}}$ to represent the total delay when connecting to the cloud server.

Based on the above, we can define the total UE delay cost under the offloading decision $\mathbf{y}(t)$ and network information $\omega(t)$ at time slot $t$ as follows:
\begin{equation}\small
\begin{aligned}  \label{UE_delay_cost}
		&\Gamma^{\rm{ue}}(\mathbf{y}(t)|\omega(t)) &\\
		&=\sum_{n\in \mathcal{N}} \sum_{m_1 \in M} y_{m_1,n}^{(\rm{s})}(t) \cdot ( T_{n,m_1 }^{(s)}(t) + D_{n,m_1 }^{(\rm{s})}(t)) \cdot f_n
\\
		&+
		\sum_{n\in \mathcal{N}} \sum_{m_2 \in \mathcal{M}} y_{m_2,n}^{(\rm{d})}(t) \cdot ( T_{n,m_2}^{(\rm{d})}(t) + D_{n,m_2}^{(\rm{d})}(t) ) \cdot f_n
\\
		&+
		\sum_{n\in \mathcal{N}} \sum_{m_1 \in \mathcal{M}}  \sum_{m_2 \in \mathcal{M}}  y_{m_1,n}^{(\rm{s})}(t) \cdot y_{m_2,n}^{(\rm{d})}(t) \cdot T_{n,m_1,m_2}^{(\rm{e})}(t) \cdot f_n
\\
		&  + \left(1-\sum_{m_1 \in M} y_{m_1,n}^{(\rm{s})}(t)  + 1- \sum_{m_2 \in M} y_{m_2,n}^{(\rm{d})}(t) \right)  \cdot T_{\rm{cld}} \cdot f_n,
\end{aligned}
\end{equation}
where the last line denotes the delay when connecting to the cloud server.

\subsubsection{Total Cost}

The total cost includes the ES deployment cost $\Gamma^{\rm{D}}( \mathbf{z} )$, determined by the ES deployment decision $\mathbf{z} $ at the beginning of the time period; the service operation cost $\Gamma^{\rm{sop}}(\mathbf{x}(t)|\omega(t))$, determined by the service placement decision $\mathbf{x}(t)$ at each time slot, and the UE delay cost $\Gamma^{\rm{ue}}(\mathbf{y}(t)|\omega(t)) $, determined by the task offloading decision $\mathbf{y}(t)$ at each time slot.
It is important to note that the ES deployment cost is calculated only once at the beginning, while the service operation cost and the UE delay cost need to be calculated at each time slot.
For convenience, we denote $\Gamma^{Q}(\mathbf{x}(t),\mathbf{y}(t)|\omega(t))$ as the weighted sum of the service operation cost and the UE delay cost at time slot $t$:
\begin{equation}\small \label{Gamma_Q}
	\begin{aligned}
		\Gamma^{Q}(\mathbf{x}(t)&,\mathbf{y}(t)|\omega(t))
		\\ & = \eta_2 \cdot \Gamma^{\rm{sop}}(\mathbf{x}(t)|\omega(t)) + \eta_3 \cdot  \Gamma^{\rm{ue}}(\mathbf{y}(t)|\omega(t)).
	\end{aligned}
\end{equation}

Then, the total system cost, denoted by $\Gamma^{\rm{ToT}}(\mathbf{z} ,\mathbf{x}(t),\mathbf{y}(t))$, can be defined as a weighted sum of the ES deployment cost and the average service operation cost and UE delay cost:
\begin{equation}\small \label{obj_total}
	\begin{aligned}
		&\Gamma^{\rm{ToT}}(\mathbf{z} ,\mathbf{x}(t),\mathbf{y}(t)) \\
		&= \eta_1 \cdot \Gamma^{\rm{D}}(\mathbf{z} )
		+ {1 \over |\T| }\sum\limits_{t=0}^{ |\T| -1} \Gamma^{Q}(\mathbf{x}(t),\mathbf{y}(t)|\omega(t))
	\end{aligned}
\end{equation}
In (\ref{Gamma_Q}) and (\ref{obj_total}), $\eta_1, \eta_2, \eta_3$ are the weights used to balance different cost components.

Based on the above, we can formulate the joint ES deployment, service placement, and computation offloading problem ($\mathbf{P0}$), aiming to minimize the long-term system-wide cost.
\begin{align} \small \label{JESCP}
	\mathbf{P0:}
	\min_{\{ \mathbf{z}, \mathbf{x}(t), \mathbf{y}(t), \forall t \in \T \}}& \notag \Gamma^{\rm{ToT}}(\mathbf{z},\mathbf{x}(t),\mathbf{y}(t)) \\  \notag
	&s.t.\quad (\ref{UE_connection_constraint}) \sim (\ref{time_avg_constraint}). \notag
\end{align}

Obviously, $\mathbf{P0}$ is a multi-timescale optimization problem,involving of the ES deployment decision in each time period, along with the service placement decision and computation offloading decision at each time-slot. However, solving $\mathbf{P0}$ is very challenging due to the following reasons.
First, the ES deployment decision $\mathbf{z}$ must be made in advance without the complete information realization, which poses a substantial challenge.
Moreover, the service placement decision $\mathbf{x}(t)$ and the computation offloading decison $\mathbf{y}(t)$ at different time slot $t \in \T$,
can be made in real time based on the information realized in each time slot, but they are coupled with each other by the long-term ES energy constraint (\ref{time_avg_constraint}) and the time-average objective function.

To capture such a temporal coupling between different decisions and information realization, we will introduce the \emph{stochastic programming} framework to transform and solve the multi-timescale problem $\mathbf{P0}$ in the next section.

\section{Problem Transformation and Solution}\label{section:problem_transformation_and_solution}

In this section, we will first introduce the stochastic programming (SP) framework to reformulate the optimization problem $\mathbf{P0}$ into a strategic-layer problem $\mathbf{P1}$ and a tactical-layer problem $\mathbf{P2}$.
Then, we will develop a Lyapunov-based algorithm to solve   problem $\mathbf{P2}$, and a Markov approximation algorithm to solve  problem $\mathbf{P1}$.

\subsection{Stochastic Programming Formulation}
\textbf{Stochastic Programming} (SP) is a framework that deals with optimization problems involving uncertainty or randomness. It is used to make optimal decisions by considering various possible scenarios and their associated probabilities.
In an SP framework, the uncertain parameters of the problem (e.g., resources at the network edge) are modeled as random variables, and the objective is to find decisions or policies that optimize expected outcomes under uncertain conditions. The key idea is to explicitly consider the inherent uncertainty in the problem and incorporate it into the optimization model.
Generally, we have a set of decisions to be made without complete information, which we call \textit{strategic-layer} decisions.
Later, complete information is revealed based on the realization of stochastic information, and the \textit{tactical-layer} decisions are taken \cite{multistage_SP}.
It is worth noting that the stochastic distribution information (e.g., the probability distribution function) is often assumed to be known a priori. \par
The general form of a stochastic programming problem is usually as follows \cite{multistage_SP}:
 \begin{equation} \label{SP-demo}
 	\begin{aligned}
 		\min_{\mathbf{x}} f_{0}(\mathbf{x})& + \mathbb{E}_{\xi}[Q^{*}(\mathbf{x},\boldsymbol{\xi})] \\
 		s.t. \quad & f_{1}(\mathbf{x}) \leq 0,
 	\end{aligned}
 \end{equation}
where
 \begin{equation} \label{Q_func}
	\begin{aligned}
		Q^{*}(\mathbf{x},\boldsymbol{\xi})=&\min_{\mathbf{y}} g_{0}(\mathbf{x,y},\boldsymbol{\xi})  \\
		&s.t. \quad g_{1}(\mathbf{x,y},\boldsymbol{\xi}) \leq 0.
	\end{aligned}
\end{equation}

In (\ref{SP-demo}) and (\ref{Q_func}), $\boldsymbol{\xi}$ represents the stochastic information, $\mathbf{x}$ is the strategic-layer decision that is made before the realization of $\boldsymbol{\xi}$,
$\mathbf{y}$ is the tactical-layer decision that is made after the realization of $\boldsymbol{\xi}$.
Moreover, $f_{0}(\mathbf{x})$ denotes the strategic-layer cost, and
$Q^{*}(\mathbf{x},\boldsymbol{\xi})$ denotes the optimal tactical-layer cost.

As for $\mathbf{P0}$, the ES deployment decision $\mathbf{z}$ is the strategic-layer decision, which is made at the beginning of each time period before the realization of information $\boldsymbol{\omega}$.
The service placement and   computation offloading decisions, i.e., $\mathbf{x}(t)$ and $\mathbf{y}(t)$, $\forall t \in \T$, are the tactical-layer decisions, which are made at each slot $t$ after the realization of the network information $\omega(t)$.
Follow the stochastic programming framework in (\ref{SP-demo}) and (\ref{Q_func}),  we can reformulate $\mathbf{P0}$ into a strategic-layer problem $\mathbf{P1}$, as shown in (\ref{ESDP}):
\begin{align} \small \label{ESDP}
	\mathbf{P1:} &\min_{\{ \mathbf{z}\}} \eta_1 \cdot \Gamma^{\rm{D}}(\mathbf{z}) + \mathbb{E}_{\omega}[Q^{*}(\mathbf{z},\boldsymbol{\omega})] \\
	s.t. &\quad (\ref{ES_deploy_constraint}), \notag
\end{align}
where
\begin{equation}  \label{opt_Q_function}
	\begin{aligned}
		Q^{*}(\mathbf{z},\boldsymbol{\omega}) \triangleq  {1 \over |\T|}\sum\limits_{t=0}^{ |\T|-1 } \Gamma^{Q}(\mathbf{x}^{*}(t),\mathbf{y}^{*}(t)|\omega(t)).
	\end{aligned}
\end{equation}

In (\ref{opt_Q_function}), $ Q^{*}(\mathbf{z},\boldsymbol{\omega}) $ denotes the
long-term average tactical-layer cost. $\mathbf{x}^{*}(t)$ and $\mathbf{y}^{*}(t)$ denote the optimal service placement and computation offloading decision, respectively.
Furthermore, we can obtain $Q^{*}(\mathbf{z},\boldsymbol{\omega})$ by solving the tactical-layer problem $\mathbf{P2}$ shown in (\ref{SPCOP}):
	\begin{align}
		\small \label{SPCOP} & \mathbf{P2:}  \min_{\{ \mathbf{x}(t), \mathbf{y}(t), \forall t \in \T \}} {1 \over |\T| }\sum\limits_{t=0}^{|\T|-1} \Gamma^{Q}(\mathbf{x}(t),\mathbf{y}(t)|\omega(t)) \\
		& ~~~~~~~~~~~~~s.t.~~~~ (\ref{UE_connection_constraint}) \sim (\ref{ES_constraint}), (\ref{ES_energy_constraint})\sim(\ref{time_avg_constraint}).  \notag                                                                                                                   \\
	\end{align}

Note that the strategic-layer problem $\mathbf{P1}$ is a multi-timescale Stochastic Mix Integer Non-Linear Programming (SMINLP) problem,
which is NP-hard.
In addition, solving $\mathbf{P1}$ requires the value of $ \mathbb{E}_{\omega}[ Q^{*}(\mathbf{z},\boldsymbol{\omega}) ] $, which need to solve the tactical-layer problem  $\mathbf{P2}$.
Thus, we will solve both problems jointly using \textit{backward induction}.
In what follows, we will first solve the tactical-layer problem $\mathbf{P2}$ in Section \ref{Section:P2}, and then solve the strategic-layer problem $\mathbf{P1}$ in Section \ref{Section:P1}.

\subsection{Lyapunov-Based Algorithm for Tactical-Layer Problem P2}\label{Section:P2}

We now solve the tactical-layer problem $\mathbf{P2}$.
It is easy to see that  $\mathbf{P2}$ is a multi-slot optimization problem with time-average objective function and time-average constraint (\ref{time_avg_constraint}).
The major challenge in solving $\mathbf{P2}$ is
to handle the long-term time-average ES energy constraint in (\ref{time_avg_constraint}).
This challenge calls for an online method that can properly make decisions without pre-knowing future information.
To address this challenge, we introduce the Lyapunov framework \cite{Neely_book} to  transform the multi-slot problem $\mathbf{P2}$ into a series of single-slot problems $\mathbf{P3}$ relying only on the local information.

\subsubsection{Virtual Queue Construction}
The key idea of the Lyapunov framework is the leveraging of virtual queues to handle long-term constraints.
As for the long-term time-average ES energy constraint (\ref{time_avg_constraint}) in $\mathbf{P2}$, we first define a \emph{virtual queue}, denoted by $\mathbf{\Theta}$, to depict the energy consumption deficit, i.e., the difference between the actual energy consumption and the long-term energy consumption budget.
Let $\mathbf{\Theta}(t)$ denote the queue backlog of $\mathbf{\Theta}$ at time slot $t$.
The dynamic evolution of such a virtual queue is shown as follows:
\begin{equation}\label{queue-update}
	\begin{aligned}
		\mathbf{\Theta}(t+1) = \max\{\mathbf{\Theta}(t)+\zeta(t)-P_{avg}, 0 \}.
	\end{aligned}
\end{equation}

According to the queue stability theorem \cite{Neely_book}, we can transform the long-term constraint (\ref{time_avg_constraint}) into the stability condition of queue.
Next, we introduce a quadratic Lyapunov function $L(\mathbf{\Theta}(t))$ to measure the stability of a queue, i.e.,
\begin{equation} \label{def_Lyapunov}
	\begin{aligned}
		L(\mathbf{\Theta}(t)) \triangleq \frac{1}{2}\mathbf{\Theta}(t)^2.
	\end{aligned}
\end{equation}

Clearly, the above Lyapunov function acts as a measure of queue stability. That is, the larger the value of the Lyapunov function, the more  unstable the queue is. Thus, a key objective in optimization is to minimize this Lyapunov function, thereby improving the queue stability.
To achieve this, we further define a drift function $\Delta(\mathbf{\Theta}(t))$ to quantify how the Lyapunov function changes over time, i.e.,
\begin{equation} \label{Lyap-drift}
	\begin{aligned}
		\Delta(\mathbf{\Theta}(t)) \triangleq L(\mathbf{\Theta}(t+1))-L(\mathbf{\Theta}(t)).
	\end{aligned}
\end{equation}

According to the Lyapunov drift theorem (Theorem 4.1 in \cite{Neely_book}), if a decision greedily minimizes the drift function $\Delta(\mathbf{\Theta}(t))$ at each time slot $t$, then the queue backlog $\mathbf{\Theta}(t)$ is gradually pushed towards a low level, which potentially maintains the stability of the queue, i.e., ensures the ES energy consumption constraint (\ref{time_avg_constraint}).

Next, we aim to address the joint queue stability and objective optimization, i.e., keeping the queue stable and meanwhile minimizing the local objective $\Gamma^{Q}(\mathbf{x}(t),\mathbf{y}(t)|\omega(t))$ at each time slot.
Following Theorem 4.2 in \cite{Neely_book}, we can~achieve this by minimizing the following \emph{drift-plus-penalty} function:
\begin{equation} \label{drift+penalty-function}
	\begin{aligned}
		\widetilde{\Delta}(\mathbf{\Theta}(t)) \triangleq \Delta(\mathbf{\Theta}(t)) + V\cdot \Gamma^{Q}(\mathbf{x}(t),\mathbf{y}(t)|\omega(t)),
	\end{aligned}
\end{equation}
where $V\geq 0$ is a tunable weight used to adjust the importance of   stability and   optimization.
Specifically, a larger $V$ is more helpful in minimizing $\Gamma^{Q}(\mathbf{x}(t),\mathbf{y}(t)|\omega(t))$, but it leads to a larger queue backlog, whereas a smaller $V$ can alleviate the queue backlog but results in a larger optimality gap.

It is noteworthy that directly minimizing the drift-plus-penalty function $\widetilde{\Delta}(\mathbf{\Theta}(t))$ is challenging due to the quadratic function in the drift $\Delta(\mathbf{\Theta}(t))$.
Thus, we focus on minimizing a specific \emph{upper-bound} of the  $\widetilde{\Delta}(\mathbf{\Theta}(t))$ to reduce complexity.
Notice that $\mathbf{\Theta}(t+1)^2 \leq (\mathbf{\Theta}(t)+\zeta(t)-P_{avg})^2$. Thus, we can derive an upper-bound of $\Delta(\mathbf{\Theta}(t))$ as follows:
\begin{equation}
	\begin{aligned}\label{que-upper-bound}
		\Delta(\mathbf{\Theta}(t))   & \leq
\frac{1}{2} (\zeta(t)-{P}_{avg})^2 + \mathbf{\Theta}(t)\cdot [\zeta(t)-{P}_{avg}]
\\
		  &\leq B +\mathbf{\Theta}(t)\cdot [\zeta(t)-{P}_{avg}],
	\end{aligned}
\end{equation}
where $\mathit{B}$ is a constant that satisfies $B \geq \frac{1}{2} (\zeta(t)-{P}_{avg})^2, \forall t\in \T $.
According to (\ref{drift+penalty-function}) and (\ref{que-upper-bound}), we can derive a potential upper-bound of $\widetilde{\Delta}(\mathbf{\Theta}(t))$ as follows:
\begin{equation} \label{drift+penalty-upper-bound}
	\begin{aligned}
		\widetilde{\Delta}(\mathbf{\Theta}(t)) \leq B & + \mathbf{\Theta}(t)\cdot [\zeta(t)-{P}_{avg}] \\
		&+ V\cdot \Gamma^{Q}(\mathbf{x}(t),\mathbf{y}(t)|\omega(t)).
	\end{aligned}
\end{equation}
Using the upper-bound derived in (\ref{drift+penalty-upper-bound}), we can propose the following single-slot optimization problem $\mathbf{P3}$:
\begin{align} \small \label{oneslot-SPCOP}
	\mathbf{P3:} \quad
	  &\min_{\{ \mathbf{x}(t), \mathbf{y}(t) \}} \mathbf{\Theta}(t)\cdot [ \zeta(t) - {P}_{avg}) ] \notag
	  \\ &~~~~~~~~~~~+ V \cdot \Gamma^{Q}(\mathbf{x}(t),\mathbf{y}(t)|\omega(t)) \\
	& s.t. \quad (\ref{UE_connection_constraint}) \sim (\ref{ES_constraint}). \notag
\end{align}
It is easy to see that the above problem $\mathbf{P3}$ relies only on the local network information, without the need of the future information.
By solving the single-slot problem $\mathbf{P3}$ at each time slot $t$, we can obtain an online tactical-layer decision, denoted as $\mathbf{x}^*(t)$ and $\mathbf{y}^*(t)$.
The above decisions will be further used to update the queue backlog according to \eqref{queue-update}, based on which we can formulate and solve the single-slot problem $\mathbf{P3}$ for the next time slot $t+1$.
Note that $\mathbf{P3}$ is an integer programming problem, which can be efficiently solved by many classical methods, such as branch-and-bound, branch-and-price, and random rounding \cite{random_rounding}. Thus, we do not delve deeply into the solution method for P3 in
this work.

We summarize such a procedure as the \underline{S}ervice \underline{P}lacement and \underline{C}omputation \underline{O}ffloading (SPCO) algorithm, illustrated in
Alg.~\ref{alg:Lyap_SPCO}.
Specifically, given an ES employment decision $\mathbf{z}$,  we observe the queue states $\mathbf{\Theta}(t)$ and the information realization $\omega(t)$ at each time slot $t\in \T$, and derive the optimal tactical-layer decisions $\mathbf{x}^*(t)$ and $\mathbf{y}^*(t)$  and the corresponding local objective value $Q^{*}(t) \triangleq \Gamma^{Q}(\mathbf{x}^{*}(t),\mathbf{y}^{*}(t)|\omega(t))$ by solving $\mathbf{P3}$.
Finally, we take the time average of ${Q}^{*}(t)$ for all time slots within the time period, i.e., $\mathbb{E}_{\omega}[Q^{*}(\mathbf{z},\boldsymbol{\omega})] = {1 \over|\T|}\sum_{t=0}^{|\T|-1} {Q}^{*}(t)$, to obtain an effective estimation of the optimal long-term time-average tactical-layer cost.




Next, we show that the SPCO algorithm not only guarantees the controllable optimality gap, but also maintain the queue stability.
Similar as Theorem 4.1 in \cite{Neely_book}, we provide the optimality gap  of the SPCO algorithm in  Theorem \ref{opt_gap_Lyap}.
\begin{theorem}[\textbf{Optimality}]\label{opt_gap_Lyap}
	The time-average tactical-layer cost achieved by the SPCO algorithm satisfy the condition:
	\begin{equation}
		\begin{aligned}
			\lim\limits_{|\T| \rightarrow \infty} \frac{1}{|\T|} \sum_{t=0}^{|\T|-1} Q^{*}(t) \leq \gamma^{\circ} + \frac{B}{V},
		\end{aligned}
	\end{equation}
where $ \gamma^\circ$ denotes the offline optimal tactical-layer cost under complete information, and $B$ is the constant   in \eqref{que-upper-bound}.
\end{theorem}
\begin{IEEEproof}
	Please refer to Appendix A in \cite{OnlineReport}.
\end{IEEEproof}

From Theorem 1, we can see that Alg.~\ref{alg:Lyap_SPCO} converges asymptotically to the optimal tactical-layer cost  $\gamma^\circ$, with a controllable approximation error bound $\O(1/V)$.

From the queue update rule (line 6 in Alg.~\ref{alg:Lyap_SPCO}), the queue backlog $ \mathbf{\Theta}(t) $ can be viewed as an indicator of the actual energy consumption of ESs.
Intuitively, a larger queue backlog implies a more significant over-consumption of energy.
Therefore, the following theorem shows the stability of queue under the SPCO algorithm.

\begin{theorem}[\textbf{Queue Stability}]\label{stable_queue}
	The time-average   queue backlog achieved by SPCO algorithm satisfy:
	\begin{equation}
		\begin{aligned}
			\lim\limits_{|\T| \rightarrow \infty} \frac{1}{|\T|}\sum_{t=0}^{|\T|-1}\mathbb{E}[ \mathbf{\Theta}(t) ]\leq \frac{B+V\cdot(\gamma_u - \gamma_l )}{\epsilon},
		\end{aligned}
	\end{equation}
	where $\gamma_u$ and $\gamma_l$ are the upper bound and lower bound of the objective function value of $\mathbf{P3}$, respectively.
\end{theorem}
\begin{IEEEproof}
	Please refer to Appendix B in \cite{OnlineReport}.
\end{IEEEproof}

From Theorem 1 and Theorem 2, we can achieve a balance between the tactical-layer cost and the ES energy consumption by adjusting the parameter $V$.
In the next section, we will further illustrate the impact of $V$ by numerical simulations.

\begin{algorithm}[t]
	\SetAlgoLined
	\KwIn{ $\mathbf{z}, \mathbf{\Theta}(0)\leftarrow 0, P_{idle}, P_{max}, P_{avg}$ and other network parameters;} 	
	\KwOut{ $\mathbf{x}^{*}(t),\mathbf{y}^{*}(t),\mathbb{E}_{\omega}[Q^{*}(\mathbf{z},\boldsymbol{\omega})], \forall t \in \T$;}	
	\For{each time slot  $\forall t \in \T$,}
	{   Observe network information $\omega(t)$; \\
        Derive  $\mathbf{x}^{*}(t), \mathbf{y}^{*}(t)$ by solving $\mathbf{P3}$; \\
        Record $Q^{*}(t) =   \Gamma^{Q}(\mathbf{x}^{*}(t),\mathbf{y}^{*}(t)|\omega(t))$; \\
        Calculate $\zeta^*(t)$ according to (\ref{ES_energy_constraint}); \\
		Update queue backlog: $\mathbf{\Theta}(t+1) = \max\{\mathbf{\Theta}(t) - P_{avg} +\zeta^*(t), 0 \}$;
	}
	Calculate $\mathbb{E}_{\omega}[Q^{*}(\mathbf{z},\boldsymbol{\omega})] \approx   {1 \over |\T| }\sum_{t=0}^{|\T|-1} {Q}^{*}(t)$; \\
	Return $\mathbb{E}_{\omega}[Q^{*}(\mathbf{z},\boldsymbol{\omega})]$,  $\mathbf{x}^{*}(t),\mathbf{y}^{*}(t), \forall t \in \T$. \\
	\caption{SPCO Algorithm}\label{alg:Lyap_SPCO}
\end{algorithm}

\subsection{Markov Approximation based Algorithm for Strategic-Layer Problem P1}
\label{Section:P1}

Now we solve Problem $\mathbf{P1}$ to obtain the ES deployment decision.
Unfortunately, $\mathbf{P1}$ is also very challenging, due to the lack of complete information.
In response to this challenge and leveraging the idea of Markov Approximation (MAP) method \cite{chenminghua_TIT2013}, we proposed the \underline{M}arkov \underline{A}pproximation based \underline{I}terative \underline{E}S deployment \underline{D}ecision (MAIED) algorithm to obtain a near-optimal solution for the ES deployment decision in $\mathbf{P1}$, with theoretically verifiable optimality gaps.

The MAP method was primarily designed for combinatorial network optimization problems.
Its core idea is to regard the network optimization problems as instances of the Max Weighted Independent Set (MWIS) problem, and aim to identify the optimal independent set that maximizes the total weight \cite{chenminghua_TIT2013}.
Following this idea, we can frame the strategic-layer problem $\mathbf{P1}$, which
involves a network consisting of a set of ES $\mathcal{M}$ and a set of configurations $\mathscr{Z}$, where each configuration in $\mathscr{Z}$ denotes a feasible choice of $\mathbf{z}$.
Building on this, we can reformulate the strategic-layer problem $\mathbf{P1}$ as the following problem $\mathbf{P4}$:
\begin{equation}   \label{ESDP-EQ}
	\begin{aligned}
	\mathbf{P4:} &\min_{ p_{\mathbf{z}} \geq 0 } \sum_{\mathbf{z} \in \mathscr{Z}} p_{\mathbf{z}}U(\mathbf{z}) \\
		&s.t. \quad \sum_{\mathbf{z}\in \mathscr{Z}} p_{\mathbf{z}}=1,
	\end{aligned}
\end{equation}
where $p_{\mathbf{z}}$ denotes the probability of choosing each feasible ES deployment decision $\mathbf{z}$ and $U(\mathbf{z})=\eta_1\cdot \Gamma^{D} + \mathbb{E}_{\omega}[Q^{*}(\mathbf{z},\boldsymbol{\omega})]$ denotes the total system cost under a particular strategy $\mathbf{z}$.

Similar as in \cite{chenminghua_TIT2013}, we can obtain a convex approximation problem $\mathbf{P5}$ for the above problem $\mathbf{P4}$, by introducing the convex log-sum-exp function. That is,
\begin{align} \small \label{MAP-ESDP}
	\mathbf{P5:}&\min_{p_{\mathbf{z}} \geq 0 } \sum_{\mathbf{z}  \in \mathscr{Z}} p_{\mathbf{z}}U(\mathbf{z}) + {1 \over \beta}\sum_{\mathbf{z}  \in \mathscr{Z}} p_{\mathbf{z}} \log{p_{\mathbf{z}}} \\
	&\quad s.t. \quad \sum_{\mathbf{z} \in \mathscr{Z}} p_{\mathbf{z}}=1, \notag
\end{align}
where $\beta$ is a positive constant that controls the approximation ratio of the system cost.
As $\beta$ approaches infinity, $\mathbf{P5}$ converges to $\mathbf{P4}$.
By using   Karush-Kuhn-Tucker (KKT) conditions, we can obtain the optimal solution to $\mathbf{P5}$:
\begin{align} \small \label{opt_p}
	p^{*}_{\mathbf{z}}={\exp \big(- \beta U(\mathbf{z}) \big) \over \sum_{\mathbf{z'} \in \mathscr{Z}} \exp \big(- \beta U(\mathbf{z'}) \big)},
\end{align}
which denotes the optimal probability of adopting each optimal ES deployment decision. \par
In order to achieve the above optimal solution, we seek to design a discrete-time Markov system with the stationary distribution in (\ref{station_markov}).
Once the Markov chain converges to a stationary distribution, the time allocation $p^{*}_{\mathbf{z}}$ for $\mathbf{z} \in \mathscr{Z}$ can be obtained \cite{chenminghua_TIT2013}.
Note that the Markov chain is required to be both time-reversible and irreducible; moreover, it traverses all feasible states under different decisions regarding ES deployment.
Therefore, the Markov chain needs to satisfy the following equation \cite{chenminghua_TIT2013}:
\begin{equation} \label{station_markov}
	p^{*}_{\mathbf{z}} \cdot p_{\mathbf{z},\mathbf{z'}} = p^{*}_{\mathbf{z'}} \cdot p_{\mathbf{z'},\mathbf{z}}, \forall \mathbf{z},\mathbf{z'}\in \mathscr{Z},
\end{equation}
where $p_{\mathbf{z},\mathbf{z'}}$ denotes the transition probability from $\mathbf{z}$ to $\mathbf{z'}$.
Note that there are many options in designing $p_{\mathbf{z},\mathbf{z'}}$. In this work, we choose the following transition probabilities:
\begin{equation}\label{trans_markov}
	p_{\mathbf{z},\mathbf{z'}}=  \bigg[1+ \alpha \exp\bigg(\beta \big(U(\mathbf{z'}) - U(\mathbf{z}) \big)\bigg)\bigg]^{-1},
\end{equation}
where $\alpha$ is a positive constant used to control the transition rate of the Markov system. Eq.~(\ref{trans_markov}) implies that $p_{\mathbf{z},\mathbf{z'}}$ is negatively related to the performance gap between the target configuration $\mathbf{z'}$ and the current  $\mathbf{z}$.
From (\ref{trans_markov}), we can further see that the system is more likely to switch to an ES deployment decision with better performance.
The parameter $\beta$ can be used to balance the exploration and exploitation, determining whether the system prioritizes discovering new solutions or refining existing ones based on past experiences.

\begin{algorithm}[t]\label{MAIED_Algorithm}
	Initialize an ES deployment decision $\mathbf{z}(0)$ and network information $\boldsymbol{\omega} $ for the first time period $l=0$\;
	Calculate $\Gamma^{D}(\mathbf{z}(0))$ by (\ref{ES_deployment_cost}); \\
	Calculate $\mathbb{E}_{\omega}[Q^{*}(\mathbf{z}(0),\boldsymbol{\omega} )]$ by Alg.~1;\\
	Set $U(\mathbf{z}(0))=\eta_1\cdot \Gamma^{D}(\mathbf{z}(0)) + \mathbb{E}_{\omega}[Q^{*}(\mathbf{z}(0),\boldsymbol{\omega} )]$;\\
	\Repeat{The last time period $|\L|$ has been reached}
	{
        Randomly generate network information $\boldsymbol{\omega}$\;
		Randomly select a feasible decision $\mathbf{z'}$\;	
		Calculate $\Gamma^{\rm{D}}( \mathbf{z'} )$ by (\ref{ES_deployment_cost})\;
		Calculate $\mathbb{E}_{\omega}[Q^{*}(\mathbf{z'},\boldsymbol{\omega})]$ by Alg.~1;\\
		Set $U(\mathbf{z'})=\eta_1\cdot \Gamma^{D}(\mathbf{z'}) + \mathbb{E}_{\omega}[Q^{*}(\mathbf{z'},\boldsymbol{\omega})]$; \\
		Calculate $ p_{\mathbf{z}(l),\mathbf{z'}}$ by (\ref{trans_markov}) ;\\
		Switch to $\mathbf{z'}$ with probability $p_{\mathbf{z}(l),\mathbf{z'}}$, and stay on $\mathbf{z}(l)$ with probality $(1-p_{\mathbf{z}(l),\mathbf{z}'})$;\\
		Set $l = l+1$; \\
		Record the current decision as $\mathbf{z}(l)$ and the corresponding system cost as $U(\mathbf{z}(l))$;
	}	
	Return the average system cost $\bar{U}=\frac{1}{|\L|} \cdot \sum_{l \in \L}U(\mathbf{z}(l))$
	\caption{MAIED Algorithm}
\end{algorithm}

%
The implementation of MAIED is described in Alg.~2.
Specifically, in each update iteration, a random ES deployment decision will be chosen for iteratively updating the state of Markov chain.
When reaching a feasible current solution $\mathbf{z}$ and the corresponding $U(\mathbf{z})$, the system record them.
To achieve a better solution, the system futher explore a target $\mathbf{z'}$ and a target objective function $U(\mathbf{z'})$.
Once the target decision $\mathbf{z'}$ shows superiority, the system accepts the target decision $\mathbf{z'}$ with probability $p_{\mathbf{z},\mathbf{z'}}$.
The iterative loop will continue until reaching a total of $|\L|$ iterations or until there is no significant improvement.
It's worth noting that, with a small value of $\beta$, the system tends to explore new solutions, this leads to an increase in convergence time.
Conversely, a large $\beta$ makes the system more likely to stay on the explored solutions.
In the next section, we will employ simulation to illustrate how the parameter $\beta$ influences system performance.

\begin{figure*}[t]
	\centering
	\subfigure[]{
		\label{tactical_layer_cost}
		\includegraphics[width=0.3\textwidth]{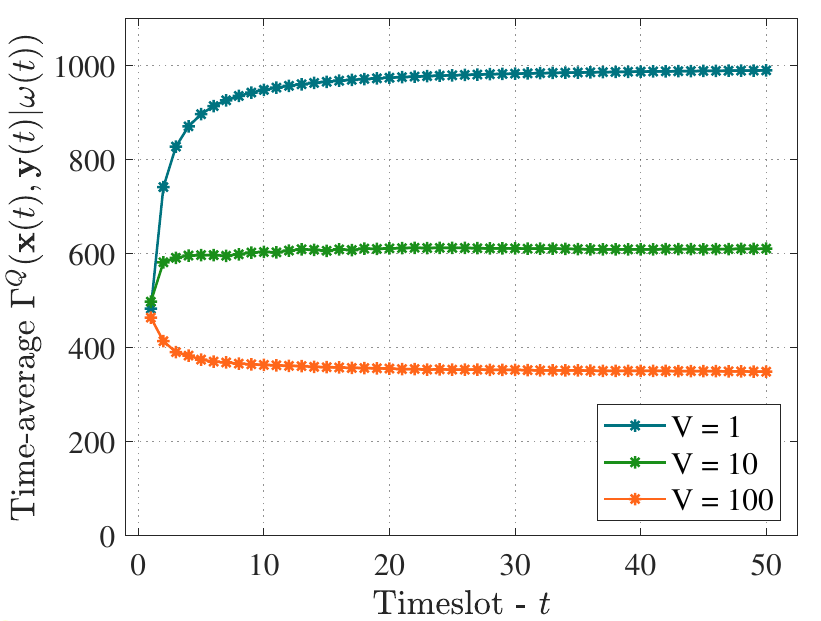}}
	\subfigure[]{
		\label{total_cost_vs_fn}
		\includegraphics[width=0.3\textwidth]{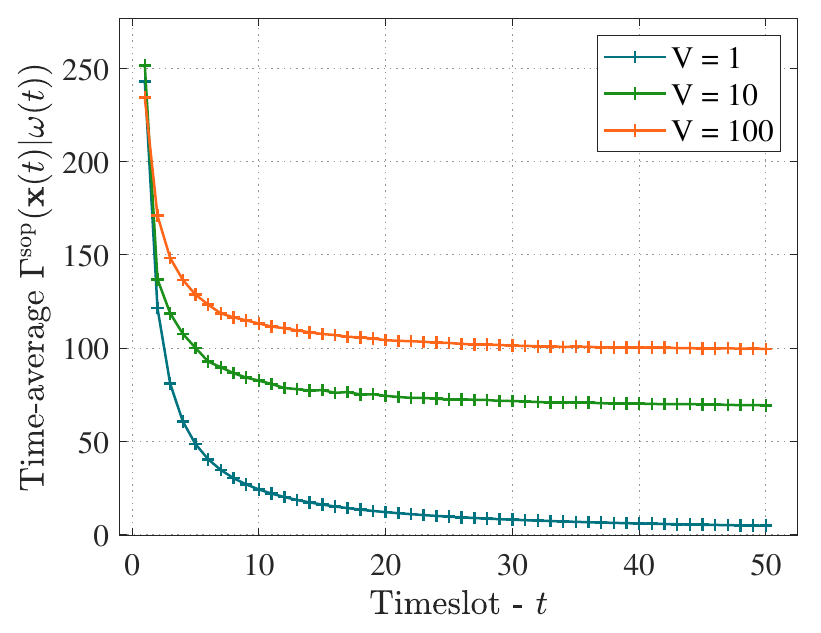}}
	\subfigure[]{
		\label{total_cost_vs_dn}
		\includegraphics[width=0.3\textwidth]{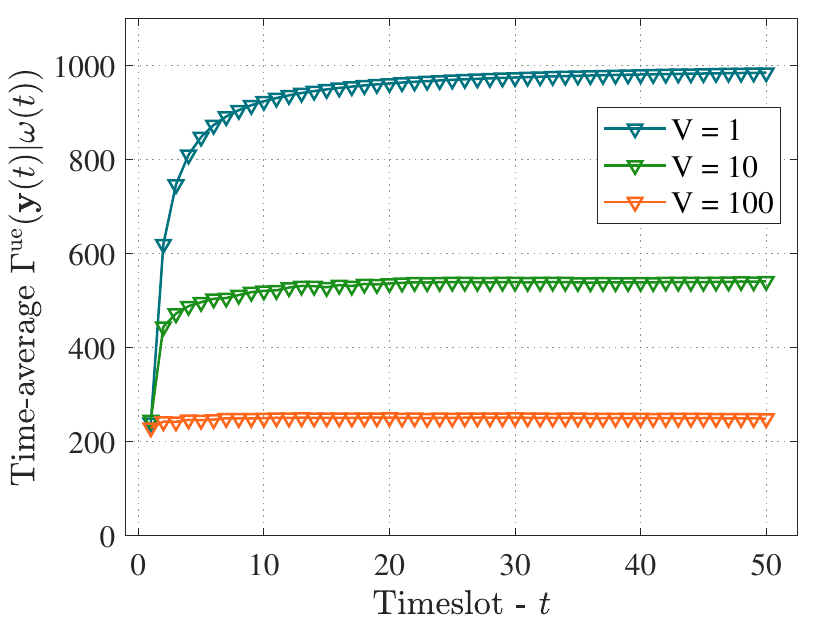}}	
	\caption{Time-average Performance of
		 (a) Tactical-layer cost
		 (b) Service operation cost
		 (c) UE delay cost.
	 }\label{Time-average performance}	
\end{figure*}

\subsection{Optimality and Complexity Analysis }
In this subsection, we will first analyze the optimality gap of the MAIED algorithm, and then examine the computational complexity of the whole solution framework.
For convinience, we refer to such a joint \underline{S}tochastic \underline{P}rogramming based \underline{J}oint \underline{E}S deployment \underline{S}ervice placement and computation \underline{O}ffloading framework as \textbf{SP-JESO}.
Clearly, the SP-JESO framework consists of the MAIED algorithm in Alg.~2 and the SPCO algorithm in Alg.~1.

\subsubsection{Optimality Gap}
The following theorem shows the optimality gap of the MAIED algorithm.
\begin{theorem}[\textbf{Optimality Gap of MAIED Algorithm}]\label{opt_gap_map}
	Let $U^{o}$ denote the optimal value of $\mathbf{P5}$, and $\bar{U}$ denote the objective value obtain by MAP. The optimality gap is bounded by:
	\begin{equation}
		\begin{aligned}
			0\leq\bar{U} - U^{o} \leq \frac{1}{\beta}\log |\mathscr{Z}|,
		\end{aligned}
	\end{equation}
where $|\mathscr{Z}|$ is the cardinal of the feasible set of ES deployment decisions of all ES.
\end{theorem}

\begin{IEEEproof}
	Please refer to Appendix C in \cite{OnlineReport}.
\end{IEEEproof}
From the above theorem, we can see that the optimality gap decreases as the parameter $\beta$ increases.
Furthermore, the optimality gap tends to grow as the feasible region of the ES deployment decision expends.
%
%
\subsubsection{Complexity Analysis}
Note that SP-JESO utilizes Alg.~2 to obtain the ES deployment decisions for each time period.
To further obtain the service placement and computation offloading decisions at each time slot, we solve $\mathbf{P3}$ in a slot-wise manner, as shown in Alg.~1.
Let the subroutine for solving $\mathbf{P3}$ at each time slot in Alg.~1 be denoted as $\mathcal{O}(1)$.
Then, the complexity of Alg.~1 is $\mathcal{O}(|\T|)$, as it requires to solve $\mathbf{P3}$ in a total of $|\T|$ times.
Accordingly, the complexity of Alg.~2 is $\mathcal{O}(|\T|\cdot |\L|)$, as it invokes Alg.~1 with $|\L|$ times.

\subsection{Practical Implications}
Now, we discuss the practical implications of the proposed solution framework.
As illustrated in Fig. \ref{fig:model}, the hybrid information decision framework introduced in this work optimally leverages available information to deliver the best performance under information constraints. This approach is particularly advantageous for emerging B5G/6G use cases, such as collaborative VR and holographic projection. For instance, in collaborative VR, the strategic deployment of edge servers (ES) ensures low-latency coverage, while tactical service placement and offloading decisions dynamically adapt to user equipment (UE) mobility and rendering workloads. Similarly, in the context of holographic projection, long-term infrastructure provisioning is planned to account for demand uncertainty, while short-term task scheduling is informed by device activity patterns. By explicitly modeling the timescale of information availability, our framework effectively balances upfront investment with runtime performance, facilitating scalability and cost-efficiency in B5G/6G edge intelligence.


\section{Simulation Results}\label{section:simulation_results}

\begin{figure*}[t]
	\centering
	\subfigure[]{
		\label{MAP_converge}
		\includegraphics[width=0.35\textwidth]{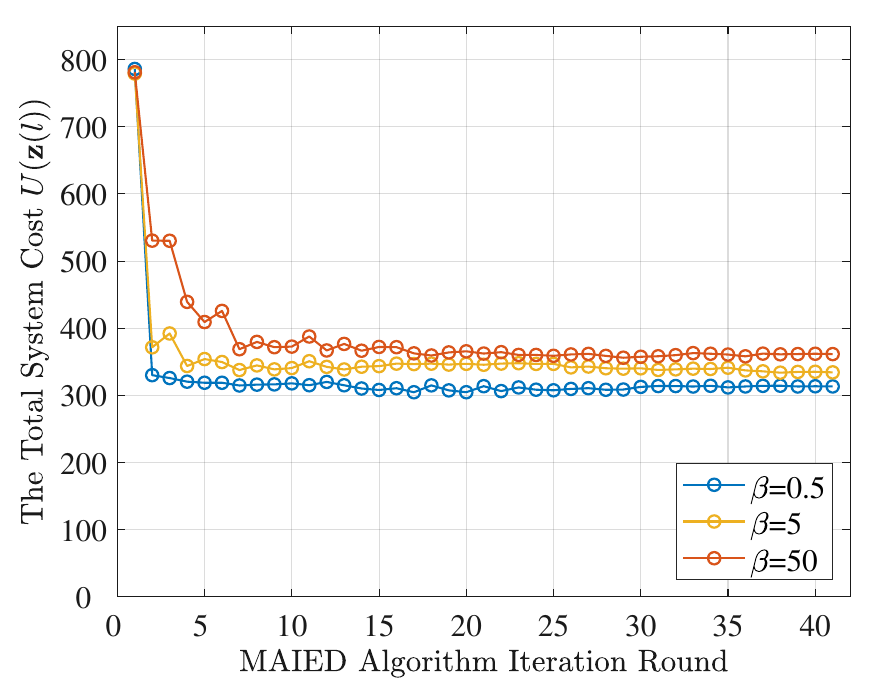}}~~~~~~~~~~~~~~~~
	\subfigure[]{
		\label{fig:fix_beta_sweep_V}
		\includegraphics[width=0.35\textwidth]{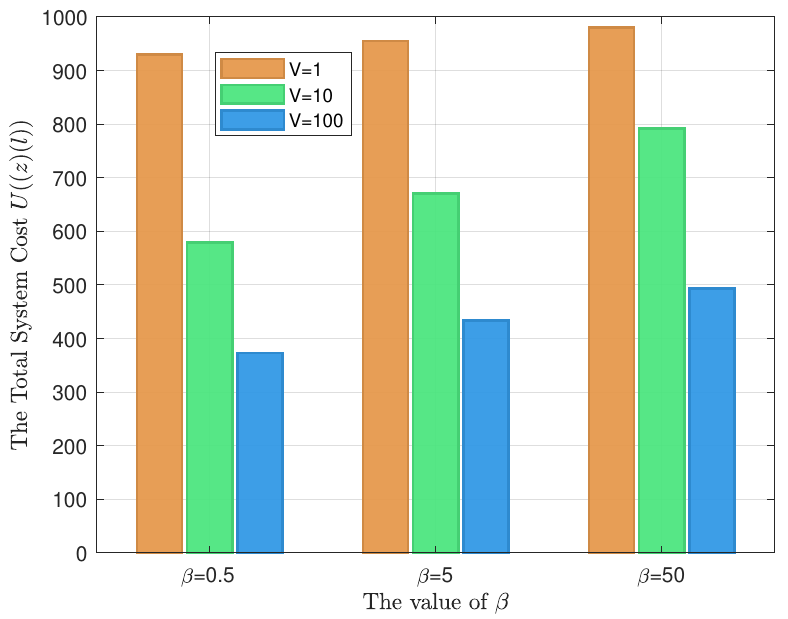}}
	\caption{Convergence properties of
		(a) The MAIED algorithm
		(b) The total system cost dynamic with $\beta-V$
	}\label{fig:beta_and_V}	
\end{figure*}
%
In this section, we illustrate the performance of SP-JESO via extensive simulations.
Specifically, we consider a 1km$\times$1km square area, where ES are distributed evenly in the area.
Meanwile, UEs follow a random walk model, where in each time slot, each UE randomly moves to a location based on a specific probability distribution.
The available resources of ES follow the normal distribution, namely, $\Phi_{m}(t) \sim \mathcal{N}(\mu,\sigma^2)$ and $\C_{m}(t) \sim \mathcal{N}(\mu,\sigma^2)$, where the mean value of  $\mu \in [50,400]$ and the standard derivation $\sigma = 5$.
The other simulation parameters are presented in Table \ref{simulation_parameter}.
Similar experiment settings have also been adopted in  \cite{Y.Gao_JIOT_2023,W.Fan_TMC2023,Y.Shi_TWC2023,R.Li_TPDS2022}.
Note that we have tested various parameter settings across different scenarios to ensure the robustness of our results.

\begin{table}[t]
	\caption{Main Simulation Parameters}
	\label{simulation_parameter}
	\vspace{-1mm}
	\begin{tabular}{l|p{12em}}
		\hline
		\textbf{Parameters} & \textbf{Default Value}  \\
		\hline
		\hline
		Storage resource of ES $\Phi_{m}(t)$ & $\Phi_{m}(t)\sim \mathcal{N}(200,25)$ \\
		Computation resource of ES $C_{m}(t)$ & $C_{m}(t)\sim \mathcal{N}(200,25)$  \\
		UE interaction frequency $f_{n}$	  &	 0.5 \\
		Wireless channel bandwidth $W$ & 2 MHz \\
		Background noise power $N_{0}$ & -174 dBm/Hz \\
		Pathloss factor $\theta$  	& 4 \\
		Lyapunov control parameter $V$ & 100 \\
		MAP control parameter $\beta$ & 50 \\
		\hline
	\end{tabular}	
\end{table}

To highlight the superiority of our proposed SP-JESO, we compare it against the following state-of-the-art solutions:
\begin{itemize}
	\item \emph{Deploying All ES} (DAE \cite{Y.Gao_JIOT_2023}): All ES are deployed regardless of the deployment budget and the realization of information.
	\item \emph{Service-Oriented ES Deployment} (SOED \cite{Y.Ma_TMC2022,Y.Shi_TWC2023}): 
The ES deployment decision is made and iteratively updated based on service operation cost information.
	\item \emph{UE-Oriented ES Deployment} (UOED \cite{Z.Bai_TMC2024}): 
The ES deployment decision is made and iteratively updated based on UE delay cost information.
\end{itemize}

In the following, we first illustrate the convergence and time-average performance of our proposed SPCO algorithm, and then compare the final results of our proposed MAIED algorithm with the above baseline solutions.

\subsection{Convergence and Performance Evaluation}
\subsubsection{Time-average Performance}

Fig. \ref{Time-average performance} illustrates  the time-average performance (i.e., the total tactical-layer cost $\Gamma^{Q}(\mathbf{x}(t),\mathbf{y}(t)|\omega(t))$, the service operation cost $\Gamma^{\mathrm{sop}}(\mathbf{x}(t)|\omega(t))$, and the UE delay cost $\Gamma^{\mathrm{ue}}(\mathbf{y}(t)|\omega(t))$) achieved in our SPCO algorithm under different values of   $V$.
From Fig.\ref{Time-average performance} (a)-(c), we can see that the time-average performance gradually converges to stable values, demonstrating the convergence of our SPCO algorithm.
Additionally, from Fig. \ref{Time-average performance} (a), we observe that a larger value of $V$ (e.g. $V=100$) leads to a smaller the tactical-layer cost $\Gamma^{Q}(\mathbf{x}(t),\mathbf{y}(t)|\omega(t))$.
This is because a larger $V$ implies that the algorithm emphasizes more on minimizing the tactical-layer cost.
From Fig.\ref{Time-average performance} (b) and (c), we can find that the service operation cost $\Gamma^{\mathrm{sop}}(\mathbf{x}(t)|\omega(t))$ increases with $V$, while the the UE delay cost $\Gamma^{\mathrm{ue}}(\mathbf{y}(t)|\omega(t))$ decreases with $V$.
This is due to the fact that a small value of
$V$ results in the absence of service placement on edge servers, forcing UEs to directly request services from the remote cloud.
Aiming to reducing the total tactical-layer cost, we will choose $V=100$ in the later simulations for  performance comparison.

\subsubsection{The Convergence of MAIED Algorithm}

Fig.~\ref{MAP_converge} illustrates the convergence behavior of the MAIED algorithm under different values of   $\beta$.
As shown in Fig.~\ref{MAP_converge}, the total system cost $U(\mathbf{z}(l))$ initially decreases and eventually stabilizes as the iterations proceed. This indicates the convergence of the MAIED algorithm for all values of $\beta\in \{0.5,5,50\}$.
Furthermore, we can observe that, a smaller value of $\beta$ (e.g., $\beta=0.5$) results in a lower system cost $U(\mathbf{z})$ in the converged state compared to a larger $\beta$ (e.g., $\beta=50$).
This can be attributed to the fact that a smaller $\beta$ encourages the system to explore a broader range of feasible ES deployment decisions.
In contrast, a larger $\beta$ tends to lead to a slightly suboptimal solution but requires fewer iterations to reach convergence.
In the subsequent simulations, we set $\beta = 5$, which strikes a good balance between solution quality and convergence speed.
Fig. \ref{fig:fix_beta_sweep_V} further presents the system cost achieved in the converged state for different values of $\beta$ and $V$.
From Fig. \ref{fig:fix_beta_sweep_V}, it is evident that, for each fixed $\beta$, the total system cost decreases as $V$ increases.
This is because a higher value of $V$ places greater emphasis on minimizing the tactical-layer cost $\Gamma^{Q}(\mathbf{x}(t),\mathbf{y}(t)|\omega(t))$, which consequently leads to a reduction in the overall system cost.

\subsection{Performance Comparison}

Now we compare the results achieved in our  SP-JESO
 with those achieved in baseline solutions.

\subsubsection{Performance vs ES Number}

As illustrated in Fig. \ref{ES_number}, the total system cost is evaluated under varying numbers of ES. The results show that as the number of ES increases, the total system cost in DAE, SP-JESO, SOED, and UOED initially decreases. This is primarily because, when the number of ES is small (e.g., $M = 1$), the UE delay cost becomes the dominant component of the total cost. Under such conditions, operators can reduce the overall system cost by deploying additional ES---within the deployment budget---to serve UEs more efficiently, thereby reducing delay-related expenses.

However, beyond a certain threshold (e.g., $M \geq  5$), adding more ES has a diminishing impact on reducing UE delay costs, causing the system cost in SP-JESO, SOED, and UOED to stabilize. In contrast, the cost in DAE starts to rise due to its failure to consider deployment budget constraints, which leads to excessive ES deployment and consequently high deployment costs.
Among all schemes, our proposed SP-JESO consistently achieves the lowest total system cost. Specifically, it outperforms DAE, SOED, and UOED by up to $56.41\%$, $33.78\%$, and
$38.93\%$, respectively.

\subsubsection{Performance vs CPU Mean Value}

Fig. \ref{ES_CPU_mean} presents the system cost under varying average CPU capacities at ES nodes. As CPU availability increases, total system cost decreases and eventually plateaus. This trend occurs because higher CPU resources enhance processing capabilities, reducing service delay and associated costs.
We can see that our proposed  SP-JESO  consistently outperforms other methods, with maximum cost reductions of $11.64\%$, $17.94\%$, and $26.88\%$, compared to DAE, SOED, and UOED, respectively.

\subsubsection{Performance vs Unit ES Deployment Cost}

Fig. \ref{ES_deploycost} shows how total system cost varies with the unit ES deployment cost $q_{m}$. As expected, the system cost increases with higher $q_{m}$ across all methods. The increase is most pronounced in DAE, which always deploys all available ES without considering cost efficiency, resulting in excessive deployment expenses. In addition, our proposed SP-JESO achieves the best performance, reducing costs by up to
$27.71\%$ over DAE, $23.29\%$ over SOED, and $26.54\%$ over UOED.

\begin{figure}[t]
	\centering
	\includegraphics[width=0.7\columnwidth]{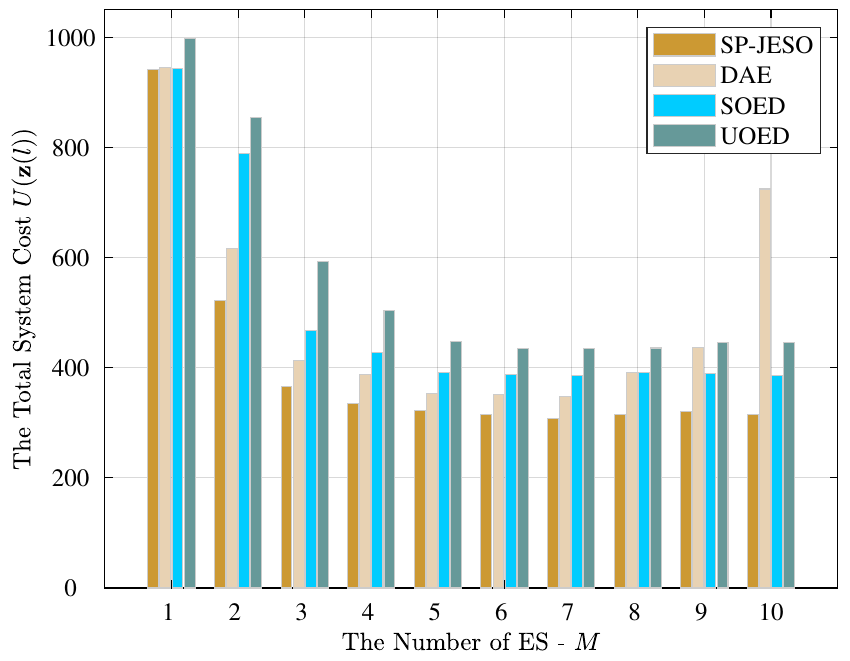}
	\caption{Total system cost vs the number of ES.}
	\label{ES_number}
\end{figure}

\begin{figure}[t]
	\centering
	\includegraphics[width=0.7\columnwidth]{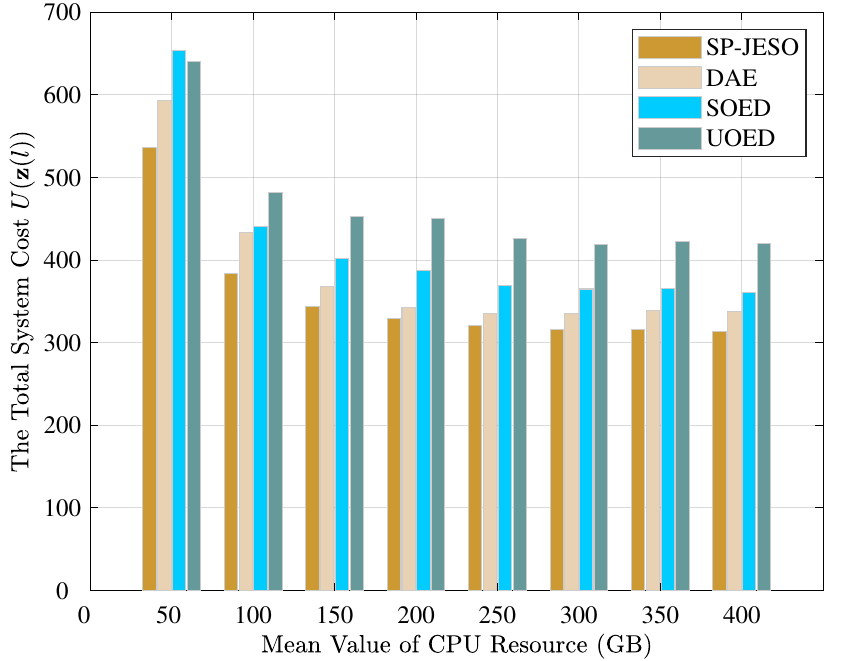}
	\caption{Total system cost vs the mean value of CPU (GHz).}
	\label{ES_CPU_mean}
\end{figure}

\begin{figure}[t]
\vspace{-1.5mm}
	\centering
	\includegraphics[width=0.7\columnwidth]{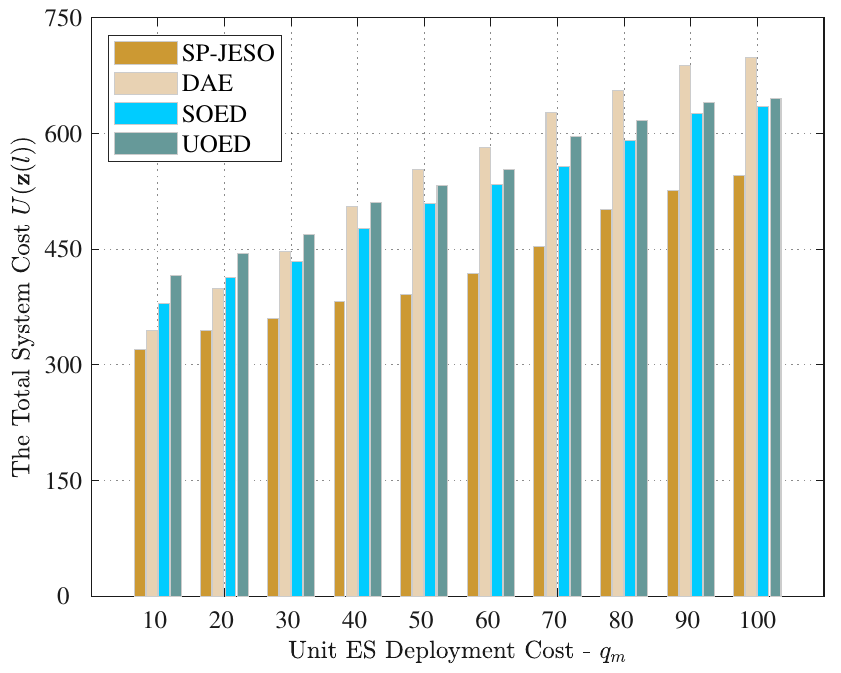}
	\caption{Total system cost vs unit ES deployment cost.}
	\label{ES_deploycost}
\end{figure}

\begin{figure}[t]
	\centering
	\includegraphics[width=0.7\columnwidth]{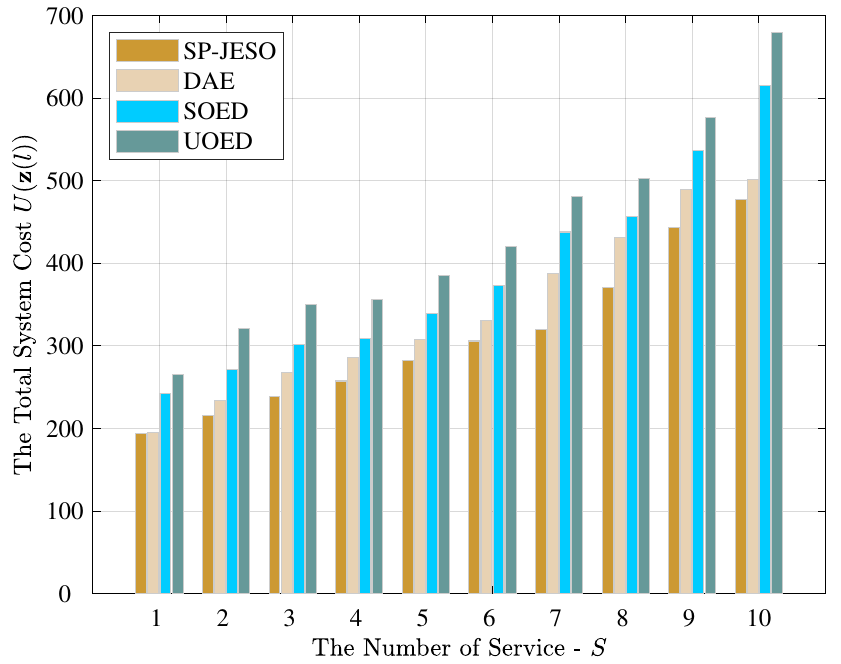}
	\caption{Total system cost vs the number of service.}
	\label{service_number}
\end{figure}

\subsubsection{Performance vs The Number of Services}

Fig. \ref{service_number} illustrates   the total system cost under different numbers of services.
From the figure, we can see that the the total system cost under all four solutions increases with the number of services.
This is due to elevated service operation costs, which scale with the number of services.
Moreover, SP-JESO consistently outperforms the other methods, achieving cost reductions of $17.41\%$,  $26.83\%$, and $33.44\%$,  compared to DAE, SOED, and UOED, respectively.

\subsubsection{Performance vs Unit Service Size}

\begin{figure}[t]
	\centering
	\includegraphics[width=0.7\columnwidth]{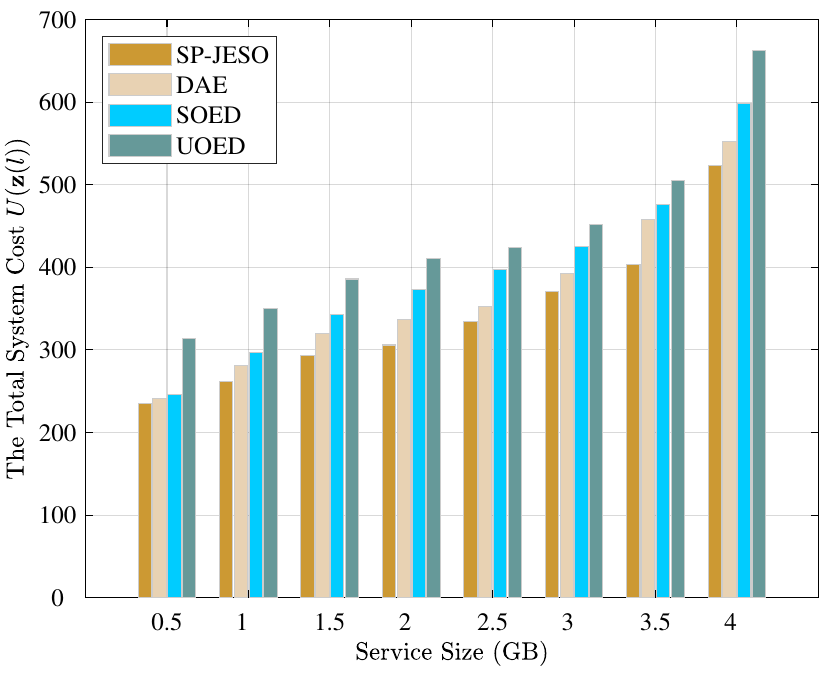}
	\caption{Total system cost vs service size.}
	\label{service_size}
\end{figure}

Fig. \ref{service_size} illustrates the total system cost under different service size.
Note taht larger services incur higher operation costs, as shown in Eqs. (\ref{service_maintenance_cost}) and (\ref{service_placement_cost}), and meanwhile require more storage space.
When ES storage is insufficient, services must be offloaded to the cloud, increasing UE delay costs.
We can find that our proposed SP-JESO continues to deliver superior performance, reducing system costs by $11.81\%$ over DAE, $18.00\%$ over SOED, and $25.46\%$ over UOED.

\begin{figure}[t]
\vspace{-2mm}
	\centering
	\includegraphics[width=0.7\columnwidth]{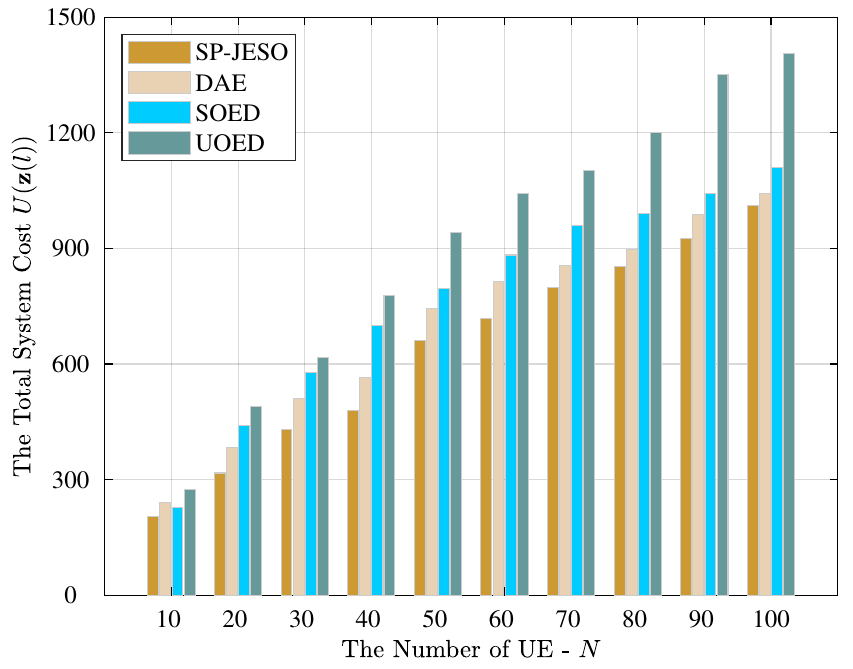}
	\caption{Total system cost vs the number of UEs.}
	\label{UE_number}
\end{figure}

\begin{figure}[t]
	\centering
	\includegraphics[width=0.7\columnwidth]{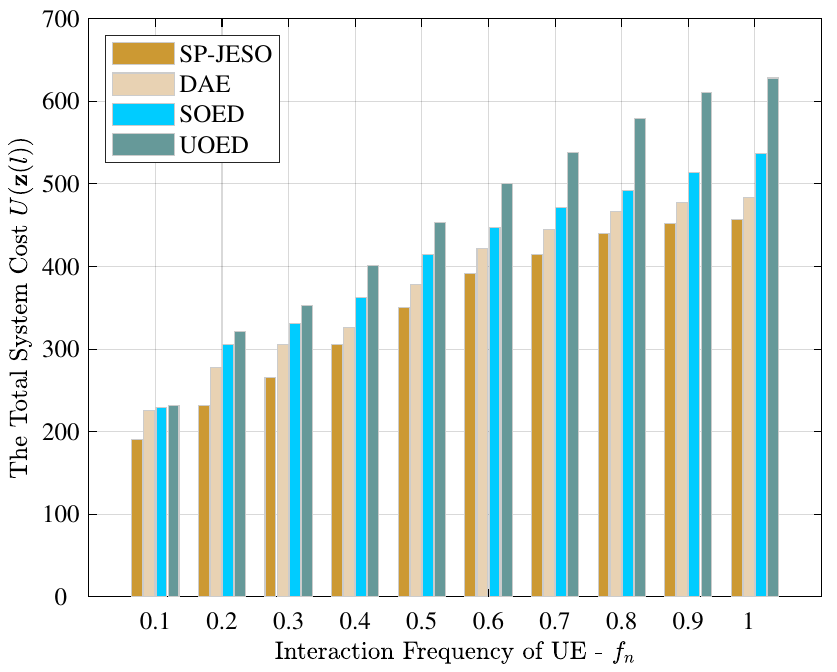}
	\caption{Total system cost vs the UE interaction frequency.}
	\label{UE_fq}
\end{figure}

\begin{figure}[t]
	\centering
	\includegraphics[width=0.7\columnwidth]{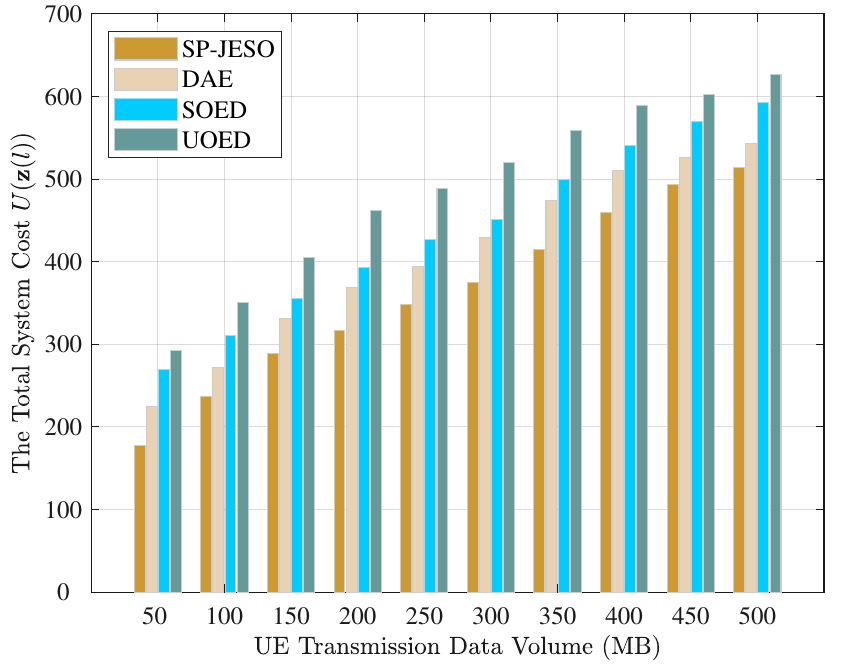}
	\caption{Total system cost vs the UE transmission data.}
	\label{UE_trans_data}
\end{figure}
\subsubsection{Performance vs UE Number}

Fig. \ref{UE_number} demonstrates   the system cost under different number of UEs.
From the figure, we can see that the the total system cost under all four solutions decreass with the number of UEs.
This is due to rising communication and computation delays and the need for additional ES deployment to handle increased load.
Moreover, our proposed SP-JESO can reduce the total system cost by up to $17.41\%$, $31.44\%$, and $38.31\%$, compared with DAE, SOED, and UOED, respectively.

\subsubsection{Performance vs UE Interaction Frequency}

Fig.~\ref{UE_fq} illustrates the total system cost for different UE interaction frequencies.
As shown in Fig.\ref{UE_fq}, a higher interaction frequency among UEs leads to an increased resource consumption and communication delays, thereby raising system costs.
Moreover, our proposed SP-JESO reduces costs by up to $16.37\%$, $24.16\%$, and $27.71\%$, compared with DAE, SOED, and UOED scheme, respectively.

\subsubsection{Performance vs UE Transmission Data}
Fig. \ref{UE_trans_data} illustrates the total system cost for different amounts of UE transmission data.
As shown in Fig. \ref{UE_trans_data}, the total system cost increases with the amount of transmission data. This is due to the fact that higher transmission data volumes lead to increased communication delay costs.
Furthermore, compared with DAE, SOED, and UOED, our proposed SP-JESO can reduce the total system cost by up to $21.32\%$, $34.32\%$, and $39.37\%$ respectively.

Fig. \ref{UE_trans_data} illustrates the total system cost for different amounts of UE transmission data. We can see that larger volumes of transmission data result in higher system costs due to increased communication delay costs. In addition, our proposed SP-JESO reduces costs by up to $16.37\%$, $24.16\%$, and $27.71\%$, compared with DAE, SOED, and UOED, respectively.

\subsubsection{Impacts of Stochastic Information}

Across the scenarios presented in Figs.\ref{ES_number} to \ref{UE_trans_data}, our proposed SP-JESO consistently achieves the lowest total system cost. The superior performance can be attributed to its ability to integrate both UE delay costs and service operation costs, while effectively considering stochastic factors and deployment budget constraints.
In contrast, SOED focuses solely on minimizing service operation costs and
UOED focuses only on UE delay costs. As a result, both approaches underperform relative to SP-JESO. Besides, SOED outperforms UOED, suggesting that service operation cost has a greater impact on the overall system cost.
The DAE method performs the worst due to its disregard for both the deployment budget and stochastic information.

\section{Conclusion}\label{section:conclusion}

In this study, we explore the joint optimization of ES deployment, service placement, and computational task offloading within a stochastic information framework. We consider a realistic scenario where the decision on ES deployment must be made in advance and remains unchanged despite the complete realization of information. In contrast, decisions on service placement and computation task offloading can be made and adjusted in real-time once full information becomes available.

To address the temporal coupling between decision-making and information realization, we introduce a SP framework. This framework encompasses a strategic layer, which is responsible for making ES deployment decisions based on incomplete stochastic information, and a tactical layer, which focuses on service placement and task offloading decisions following complete information realization.
To address the challenges arising from the different timescales of the decisions at the two layers, we propose a multi-timescale SP framework. This includes a large timescale period for strategic-layer decision-making and a smaller timescale slot for tactical-layer decision-making. To solve the tactical-layer problem at each time slot, we developed a Lyapunov-based algorithm. Additionally, a Markov approximation algorithm was designed to solve the strategic-layer problem for each time period.
Simulation results demonstrate that our proposed solution significantly outperforms existing benchmarks that do not account for the coupling between decision-making and information realization, achieving up to   $56\%$ reduction in total system cost.


\bibliographystyle{unsrt}
\bibliographystyle{IEEEtran} 




\begin{IEEEbiography}[{\includegraphics[width=1in,height=1.25in,clip,keepaspectratio]{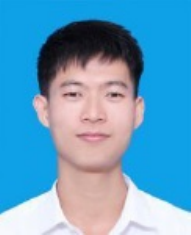}}]{Jiaqi Wu} (Student Member, IEEE) is currently a PhD student at the School of Electronics and Information Engineering, Harbin Institute of Technology, Shenzhen, China. He received the Master degree from the Guangxi University, China, in 2020.
His main research interests are in the interdisciplinary area between network optimization and machine learning, with particular focuses on the deep reinforcement learning and mobile edge computing.
\end{IEEEbiography}

\begin{IEEEbiography}[{\includegraphics[width=1in,clip,keepaspectratio]{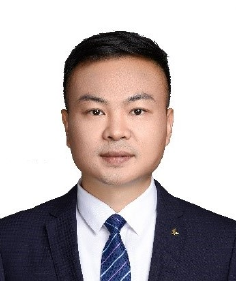}}]{Bin Cao}
(Member, IEEE) received the Ph.D. degree in Information and Communication Engineering from Harbin Institute of Technology, Shenzhen, China, in 2013.
He is currently an Associate Professor with the School of Electronics and Information
Engineering, Harbin Institute of Technology, Shenzhen, China.
From 2010 to 2012, he was a Visiting Scholar with the University of Waterloo, Canada.
His research interests include signal processing for wireless communications, cognitive radio networking, and resource allocation for wireless networks.
\end{IEEEbiography}

\begin{IEEEbiography}[{\includegraphics[width=1in,height=1.25in,clip,keepaspectratio]{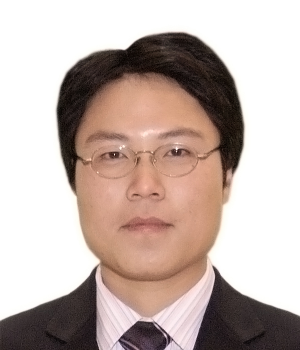}}]{Lin Gao} (Senior Member, IEEE) is a Professor at the School of Electronics and Information Engineering, Harbin Institute of Technology, Shenzhen, China. He received the Ph.D. degree in Electronic Engineering from Shanghai Jiao Tong University, Shanghai, China, in 2010. His main research interests are in the interdisciplinary area between optimization, game theory, and artificial intelligence, with particular focuses on reinforcement learning, federated learning, crowd/edge intelligence, mobile edge computing,  and cognitive networking. He is the co-recipient of 5 Best Paper Awards from leading conference proceedings on wireless communications and networking. He received the IEEE ComSoc Asia-Pacific Outstanding Young Researcher Award in 2016.
\end{IEEEbiography}

\end{document}